\documentclass[11pt]{article}
\usepackage{amsmath,amssymb,amsthm,amsxtra,overpic,bbm,bm,epsfig,ulem,color,multirow}
\usepackage{braket}
\usepackage{makecell}

\begin{document}

\begin{center}
{\Large Renormalization group running triggered tri-resonant leptogenesis \\
 within neutrino flavor-symmetry models }
\end{center}

\vspace{0.05cm}

\begin{center}
{\bf Zhi-Cheng Liu$^1$, Zhen-hua Zhao$^2$\footnote{Corresponding author: zhaozhenhua@lnnu.edu.cn}} \\
{ $^1$ School of Materials Science and Engineering, Anyang Institute of Technology, \\ Anyang 455000, China \\
$^2$ Department of Physics, Liaoning Normal University, Dalian 116029, China
}
\end{center}

\vspace{0.2cm}

\begin{abstract}
In this paper, we have studied the consequences of some representative neutrino flavor-symmetry models for tri-resonant leptogenesis (which is realized by having three nearly degenerate right-handed neutrinos). To be specific, we have considered a neutrino flavor-symmetry model realizing the TM1 mixing and modular ${\rm A}^{}_4$, ${\rm S}^{}_4$ and ${\rm A}^{}_5$ symmetry models that have a right-handed neutrino mass matrix $M^{}_{\rm R}$ as shown in Eq.~(\ref{6}) which gives three exactly degenerate right-handed neutrino masses and consequently prohibits the leptogenesis mechanism to work successfully. For these models, we study the scenario that the desired right-handed neutrino mass splittings for leptogenesis to work are generated from the renormalization-group running effects. In such a scenario, we explore the parameter space that allows for the reproduction of the observed baryon-antibaryon asymmetry of the Universe.
\end{abstract}

\newpage

\section{Introduction}

As we know, the discovery of neutrino oscillations reveals that neutrinos are massive and their flavor eigenstates $\nu^{}_\alpha$ (for $\alpha =e, \mu, \tau$) are certain superpositions of their mass eigenstates $\nu^{}_i$ (for $i =1, 2, 3$) possessing definite masses $m^{}_i$: $\nu^{}_\alpha = \sum^{}_i U^{}_{\alpha i} \nu^{}_i$ with $U^{}_{\alpha i}$ being the $\alpha i$-element of the neutrino mixing matrix $U$ \cite{xing}. In the standard form, $U$ is expressed in terms of three mixing angles $\theta^{}_{ij}$ (for $ij=12, 13, 23$), one Dirac CP phase $\delta$ and two Majorana CP phases $\alpha^{}_{21}$ and $\alpha^{}_{31}$ as
\begin{eqnarray}
U  = \left( \begin{matrix}
c^{}_{12} c^{}_{13} & s^{}_{12} c^{}_{13} & s^{}_{13} e^{-{\rm i} \delta} \cr
-s^{}_{12} c^{}_{23} - c^{}_{12} s^{}_{23} s^{}_{13} e^{{\rm i} \delta}
& c^{}_{12} c^{}_{23} - s^{}_{12} s^{}_{23} s^{}_{13} e^{{\rm i} \delta}  & s^{}_{23} c^{}_{13} \cr
s^{}_{12} s^{}_{23} - c^{}_{12} c^{}_{23} s^{}_{13} e^{{\rm i} \delta}
& -c^{}_{12} s^{}_{23} - s^{}_{12} c^{}_{23} s^{}_{13} e^{{\rm i} \delta} & c^{}_{23}c^{}_{13}
\end{matrix} \right) \left( \begin{matrix}
1 &  & \cr
& e^{{\rm i}\alpha^{}_{21}}  & \cr
&  & e^{{\rm i}\alpha^{}_{31}}
\end{matrix} \right) \;,
\label{1}
\end{eqnarray}
where the abbreviations $c^{}_{ij} = \cos \theta^{}_{ij}$ and $s^{}_{ij} = \sin \theta^{}_{ij}$ have been employed.

Nowadays, the neutrino mixing angles and neutrino mass squared differences $\Delta m^2_{ij} \equiv m^2_i - m^2_j$ have been measured to good degrees of accuracy, and there is also a preliminary result for $\delta$ (but with a large uncertainty). Several research groups have performed global analyses of the existing neutrino oscillation data to extract the values of these parameters \cite{global}. As is known, the sign of $\Delta m^2_{31}$ remains undetermined, thereby allowing for two possible neutrino mass orderings: the normal ordering (NO) $m^{}_1 < m^{}_2 < m^{}_3$ and inverted ordering (IO) $m^{}_3 < m^{}_1 < m^{}_2$. On the other hand, neutrino oscillations are completely insensitive to the absolute neutrino mass scale and the Majorana CP phases. Their values can only be inferred from certain non-oscillatory experiments such as the neutrinoless double beta decay experiments \cite{0nbb}. But so far there has not been any lower bound on the lightest neutrino mass, nor any constraint on the Majorana CP phases.

For the neutrino mixing angles, it is known that $\theta^{}_{12}$ and $\theta^{}_{23}$ are close to some special values (i.e., $\sin^2 \theta^{}_{12} \sim 1/3$ and $\sin^2 \theta^{}_{23} \sim 1/2$), while $\theta^{}_{13}$ is relatively small. In fact, before the true value of $\theta^{}_{13}$ was experimentally pinned down, it had been widely believed to be vanishingly small. For the ideal case of $\sin^2 \theta^{}_{12} = 1/3$, $\sin^2 \theta^{}_{23} = 1/2$ and $\theta^{}_{13} =0$, the neutrino mixing matrix takes a very simple form as
\begin{eqnarray}
U^{}_{\rm TBM}= \displaystyle \frac{1}{\sqrt 6} \left( \begin{array}{ccc}
2 & \sqrt{2} & 0 \cr
1 & - \sqrt{2}  & -\sqrt{3}  \cr
1 & - \sqrt{2}  & \sqrt{3} \cr
\end{array} \right)  \;,
\label{2}
\end{eqnarray}
which is referred to as the tribimaximal (TBM) mixing \cite{TB}. This particular mixing has inspired intensive model-building studies with the employment of certain discrete non-Abelian symmetries (e.g., ${\rm A}^{}_4$ and ${\rm S}^{}_4$) \cite{FS}.
However, the observation of a relatively large $\theta^{}_{13}$ (compared to $0$) motivates us to make corrections for the TBM mixing. In this connection, an economical and natural choice is to retain its first or second column while modifying the other two columns within the unitary constraints, thus yielding the first or second trimaximal (TM1 or TM2) mixing \cite{TM}
\begin{eqnarray}
U^{}_{\rm TM1}=  \displaystyle \frac{1}{\sqrt 6} \left( \begin{array}{ccc}
2 & \cdot & \cdot \cr
1 & \cdot & \cdot \cr
1 & \cdot & \cdot \cr
\end{array} \right)  \;, \hspace{1cm}
U^{}_{\rm TM2}=  \displaystyle \frac{1}{\sqrt 3} \left( \begin{array}{ccc}
\cdot & 1 & \cdot \cr
\cdot & -1 &  \cdot \cr
\cdot &  -1 & \cdot \cr
\end{array} \right)  \;,
\label{3}
\end{eqnarray}
where the dot signs denote the unspecified elements.
Due to their simplicity and interesting consequences, the TM1 and TM2 mixings have attracted a lot of attention and many neutrino flavor-symmetry models have been formulated to realize them \cite{FS}.

An obvious shortcoming of the conventional flavor-symmetry models is that many flavon fields need to be introduced in order to achieve a desired breaking pattern of the flavor symmetry, and the scalar potential of these flavon fields needs a complicated symmetric shaping. Moreover, higher-dimensional operators with more flavon insertions and unknown coefficients are often present, affecting the predictive power of the models. In recent years, the idea of modular invariance which provides substantial simplifications to these problems has been proposed and become increasingly popular \cite{modular, modularR}. In the modular-invariance approach, the flavor symmetry is broken by the vacuum expectation value (VEV) of a single scalar field---the so-called modulus, while the flavon fields are not necessary. And the Yukawa couplings and fermion mass terms are certain special functions of the modulus called modular forms that have particular properties under the modular group transformations.
Moreover, all higher-dimensional operators are unambiguously determined in the limit of unbroken supersymmetry. As a result, flavor models with modular invariance depend on fewer parameters, enhancing their predictive power.

As for the neutrino masses, one of the most popular and natural ways of generating them is the type-I seesaw model \cite{seesaw} in which three heavy right-handed neutrinos $N^{}_I$ ($I=1, 2, 3$) are introduced into the Standard Model (SM). First of all, $N^{}_I$ can constitute the Yukawa coupling operators together with the lepton doublets $L^{}_\alpha$ and the Higgs doublet $H$: $Y^{}_{I \alpha} \overline {N^{}_I } H L^{}_\alpha$ with $Y^{}_{I \alpha}$ being the $I \alpha$-element of the Yukawa coupling matrix $Y$. These operators will generate the Dirac neutrino mass terms $(M^{}_{\rm D})^{}_{I \alpha}= Y^{}_{I \alpha} v$ (here $(M^{}_{\rm D})^{}_{I \alpha}$ is the $I \alpha$-element of the Dirac neutrino mass matrix $M^{}_{\rm D}$) after the neutral component of $H$ acquires a nonzero VEV $v = 174$ GeV. In addition, $N^{}_I$ themselves can also have the Majorana mass terms $\overline{N^{}_I} (M^{}_{\rm R})^{}_{IJ} N^{c}_J$ (here $(M^{}_{\rm R})^{}_{IJ}$ is the $IJ$-element of the right-handed neutrino mass matrix $M^{}_{\rm R}$).
Then, under the seesaw condition $M^{}_{\rm R} \gg M^{}_{\rm D}$, integrating the right-handed neutrinos out yields an effective Majorana mass matrix for three light neutrinos as
\begin{eqnarray}
M^{}_{\nu} = - M^{T}_{\rm D} M^{-1}_{\rm R} M^{}_{\rm D} \;.
\label{4}
\end{eqnarray}
Here we would like to emphasize that throughout the paper the fermion mass matrices are written in the right-left convention which is more conveniently used in the supersymmetric setup as applicable to our study.

Remarkably, the seesaw model also provides an attractive explanation (known as the leptogenesis mechanism \cite{leptogenesis, Lreview}) for the baryon-antibaryon asymmetry of the Universe \cite{planck}
\begin{eqnarray}
\eta^{}_{\rm B} \equiv \frac{n^{}_{\rm B}-n^{}_{\rm \bar B}}{n^{}_\gamma} \simeq (6.11 \pm 0.04) \times 10^{-10}  \;,
\label{5}
\end{eqnarray}
where $n^{}_{\rm B}$ ($n^{}_{\rm \bar B}$) denotes the baryon (antibaryon) number density and $n^{}_\gamma$ the photon number density. The leptogenesis mechanism works in a way as follows: a lepton-antilepton asymmetry $\eta^{}_{\rm L} \equiv (n^{}_{\rm L} - n^{}_{\rm \bar L})/n^{}_\gamma$ is firstly generated from the out-of-equilibrium and CP-violating decays of the right-handed neutrinos and then partly converted into the baryon-antibaryon asymmetry via the sphaleron processes: $\eta^{}_{\rm B} \simeq - c\eta^{}_{\rm L}$ with $c = 28/79$ or $8/23$ in the SM or MSSM (Minimal Supersymmetric Standard Model) framework.

Unfortunately, for some typical neutrino flavor-symmetry models (see those given in sections~3-6), the right-handed neutrino mass matrix takes a form as
\begin{eqnarray}
M^{}_{\rm R} = M^{}_0 \begin{pmatrix}
	 1 & 0 & 0  \\ 0 & 0 & 1  \\ 0 & 1  & 0
\end{pmatrix} \;,
\label{6}
\end{eqnarray}
which gives three exactly degenerate right-handed neutrino masses (i.e., $M^{}_0$). In this case, the leptogenesis mechanism is prohibited to work successfully (to be specific, there would no CP asymmetries for the decays of right-handed neutrinos). But just as a coin has two sides, this kind of models can serve as a unique basis for realizations of the tri-resonant leptogenesis scenario \cite{resonant}: if the right-handed neutrino masses receive certain corrections so that their degeneracy is lifted to an appropriate extent, then the tri-resonant leptogenesis scenario will be naturally realized.

In this paper, within this kind of neutrino flavor-symmetry models, we study the scenario that  the desired right-handed neutrino mass splittings for leptogenesis to work are generated from the renormalization-group equation (RGE) effects \cite{rgeL}: if the energy scale $\Lambda^{}_{\rm FS}$ of the flavor-symmetry physics that shapes the special textures of neutrino mass matrices is much higher than the right-handed neutrino mass scale $M^{}_0$ (throughout this paper, $M^{}_0$ will be used to denote the common masses of three nearly degenerate right-handed neutrinos when their splitting is not important) where leptogenesis takes place, then the RGE effect between the scales of $\Lambda^{}_{\rm FS}$ and $M^{}_0$ may induce some mass splittings among the three right-handed neutrinos (see section~2.2). Such a correction of $M^{}_{\rm R}$ has the following two merits: 1) this effect is spontaneous, provided that there is a considerable gap between the flavor-symmetry scale and the right-handed neutrino mass scale; 2) this effect is minimal, in the sense that it does not need to introduce additional parameters.

The remaining parts of this paper are organized as follows. In section~2 we will first give some basic formulas about leptogenesis and RGE effects that will be used in our study. Then, we perform the study for four representative neutrino flavor-symmetry models which contain three degenerate right-handed neutrinos \footnote{For a similar study about the $\Delta(3n^2)$ and $\Delta(6n^2)$ neutrino flavor-symmetry models, see Ref.~\cite{drewes}.}. We successively study a neutrino flavor-symmetry model realizing the TM1 mixing in section~3,
modular ${\rm A}^{}_4$ symmetry model in section~4, modular ${\rm S}^{}_4$ symmetry model in section~5, and modular ${\rm A}^{}_5$ symmetry model in section~6. In section~7, we summarize our main results.

\section{Some formulas about leptogenesis and RGE effects}

\subsection{Some formulas about leptogenesis}

In the resonant leptogenesis scenario, the CP asymmetries for the decay processes of right-handed neutrinos will be greatly enhanced due to nearly degenerate masses of the right-handed neutrinos.
The resonance effects can be encoded in the effective neutrino Yukawa couplings as \cite{resonant}
\begin{eqnarray}
	&&\hspace{-2cm} \left( \bar{Y}^{}_{+} \right)^{}_{I \alpha} = Y_{I \alpha } - {\rm i} V^{}_{I \alpha}
	+ {\rm i} \sum_{J, K=1}^{3} \left| \epsilon^{}_{IJK} \right|  Y^{}_{J \alpha } \nonumber \\
	&& \hspace{-0.2cm} \times
	\frac{ M^{}_{I}  \left( \mathcal{M}^{}_{IIJ } + \mathcal{M}^{}_{JJI} \right) + {\rm i} R^{}_{IK}
		\left[ \mathcal{M}^{}_{IKJ } \left( \mathcal{M}^{}_{IIK } + \mathcal{M}^{}_{KKI } \right) + \mathcal{M}^{}_{JJK } \left( \mathcal{M}^{}_{IKI } + \mathcal{M}^{}_{KIK} \right) \right]
	}
	{ M^{2}_{I} - M^{2}_{J} - 2 {\rm i} M^{2}_{I} A^{}_{JJ}
		- 2 {\rm i} {\rm Im}( R^{}_{IK}) \left[ M^{2}_{I} \left| A^{}_{JK}\right|^2 + M^{}_{J} M^{}_{K} {\rm Re} ( A^2_{JK}) \right]  } \;,
	\label{2.1}
\end{eqnarray}
where $\epsilon^{}_{IJK} $ is the anti-symmetric Levi-Civita symbol, $\mathcal{M}^{}_{IJK} \equiv M^{}_{I} A^{}_{JK}$ and
\begin{eqnarray}
	&& A^{}_{IJ} = \frac{ \left( Y Y^{\dagger} \right)^*_{IJ} }{16 \pi} \;, \hspace{1cm}
	V^{}_{I \alpha} =\sum^{}_{\beta =e, \mu, \tau} \sum^{}_{K \neq I} \frac{Y^{*}_{I \beta} Y^{}_{K \beta} Y^{}_{K \alpha} }{16 \pi}
	f\left( \frac{M^2_{K}}{M^2_{I}} \right) \;, \nonumber \\
	&& R^{}_{IJ} = \frac{ M^{2}_{I} }{ M^{2}_{I} - M^{2}_{J} - 2 {\rm i} M^{2}_{I} A^{}_{JJ} } \;,
	\label{2.2}
\end{eqnarray}
with $f(x) = \sqrt{x} [ 1 - (1+x) \ln (1+1/x)]$ being the loop function.
And the corresponding CP-conjugate effective Yukawa couplings $\left( \bar{Y}^{}_{-} \right)^{}_{I \alpha}$ can be obtained by simply making the replacement $Y^{}_{I \alpha} \to Y^{*}_{I \alpha}$ in the above expressions. In terms of these effective Yukawa couplings, the decay widths of right-handed neutrinos can be expressed as
\begin{eqnarray}
\Gamma^{}_I \equiv \sum^{}_{\alpha} \Gamma \left( N^{}_I \to L^{}_{\alpha} + H \right)
= \frac{ M^{}_{I} }{16\pi} \left| \left( \bar{Y}^{}_{+} \bar{Y}^{\dagger}_{+} \right)^{}_{I I} \right|^2 \;, \quad
\bar{\Gamma}^{}_I  \equiv \sum^{}_{\alpha} \Gamma \left( N^{}_I \to \overline{L}^{}_\alpha + \overline{H} \right)
= \frac{ M^{}_{I} }{16\pi} \left| \left( \bar{Y}^{}_{-} \bar{Y}^{\dagger}_{-} \right)^{}_{I I} \right|^2 \;,
	\label{2.3}
\end{eqnarray}
Finally, the CP-asymmetry matrix element $\varepsilon^{\left(I\right)}_{\alpha \beta}$ describing the asymmetries generated from the decays of $N^{}_I$ can be obtained as
\begin{eqnarray}
	\varepsilon^{\left(I\right)}_{\alpha \beta}
	= \frac{ \bar{\mathcal{P}}^{\left(I\right) }_{\alpha \beta} \bar{\Gamma}^{}_I - \mathcal{P}^{\left(I\right)}_{\alpha \beta} \Gamma^{}_I}{ \Gamma^{}_I + \bar{\Gamma}^{}_I } \;.
	\label{2.4}
\end{eqnarray}
Here the two projectors $\mathcal{P}^{\left(I\right)}_{\alpha \beta}$ and $\bar{\mathcal{P}}^{\left(I\right) }_{\alpha \beta}$ are given by
\begin{eqnarray}
	\mathcal{P}^{\left(I\right)}_{\alpha \beta} \equiv C^{}_{\alpha I} C^{*}_{\beta I} \;,\qquad
	\bar{\mathcal{P}}^{\left(I\right) }_{\alpha \beta} \equiv \bar{C}^{}_{\alpha I} \bar{C}^{*}_{\beta I} \;,
	\label{2.5}
\end{eqnarray}
with
\begin{eqnarray}
	C^{}_{I \alpha } = \frac{ \left( \bar{Y}^{}_{+} \right)^{}_{I \alpha} }{ \sqrt{\left(\bar{Y}^{}_{+} \bar{Y}^\dagger_{+}\right)^{}_{II}} } \;, \qquad
	\bar{C}^{}_{I \alpha } = \frac{ \left( \bar{Y}^{}_{-} \right)^{}_{I \alpha} }{ \sqrt{\left(\bar{Y}^{}_{-} \bar{Y}^\dagger_{-}\right)^{}_{II}} } \;.
\end{eqnarray}
Neglecting CP-violating corrections, $C^{}_{\alpha I}$ and $\bar{C}^{}_{\alpha I}$ reduce to
\begin{eqnarray}
C^{0}_{I \alpha } = \frac{ Y^{}_{I \alpha} }{ \sqrt{\left(Y Y^\dagger_{ }\right)^{}_{II}} } \;.
	\label{2.6}
\end{eqnarray}
Note that the diagonal terms of the CP-asymmetry matrix correspond to the flavoured CP asymmetries
(i.e., $\varepsilon^{}_{I\alpha} = \varepsilon^{\left(I\right)}_{\alpha \alpha}$), and the total CP asymmetries are the trace of the CP-asymmetry matrix (i.e., $\varepsilon^{}_{I} = \sum^{}_{\alpha } \varepsilon^{\left(I\right)}_{\alpha \alpha}$).

In order to generate a nonzero lepton asymmetry in the Universe, one needs not only the CP asymmetries, but also a departure from thermal equilibrium. In most cases, classical Boltzmann equations are sufficient for the calculation of the final asymmetry \cite{Lreview}. However, when lepton flavour effects are taken into account \cite{flavor}, different sets of classical Boltzmann equations apply depending whether the asymmetry is generated in the unflavored regime (where none of the charged-lepton Yukawa interactions has entered into equilibrium), in the two-flavour regime (where only the tauonic Yukawa interaction has entered into equilibrium), or in the three-flavour regime (where both the tauonic and muonic Yukawa interactions have entered into equilibrium). In particular, classical Boltzmann equations would fail in reproducing the correct result in the transition regimes between any two different flavor regimes. As our study aims to explore the full parameter space of the right-handed neutrino masses, we will employ the density matrix formalism \cite{Blanchet:2011xq} which applies to different flavor regimes as well as the transition regimes. In this formalism the evolutions of the right-handed neutrino number density $n^{}_{N^{}_I}$ and lepton asymmetry number density $n^{}_{\alpha \beta}$ as functions of the temperature (via $z \equiv M^{}_1/T$) can be tracked by the following equations \cite{Blanchet:2011xq}
\begin{eqnarray}
	\frac{ {\rm d} n^{}_{N^{}_I} }{{\rm d} z} &=& - D^{}_I \left( n^{}_{N^{}_I} - n^{\rm eq}_{N^{}_I} \right) \;,
\nonumber	\\ 
	\frac{ {\rm d} n^{}_{ee} }{{\rm d} z} &=& \sum_{I=1}^{3} \left\{ \varepsilon^{\left(I\right)}_{ee} D^{}_I \left( n^{}_{N^{}_I} - n^{\rm eq}_{N^{}_I} \right)
	- W^{}_I \left[ \left|C^{0}_{Ie}\right|^2 n^{}_{ee} + {\rm Re} \left( C^{0}_{Ie} C^{0}_{I\mu} n^{*}_{\mu e} +  C^{0}_{Ie} C^{0}_{I\tau} n^{*}_{\tau e} \right) \right]
	\right\}  \;,
\nonumber	\\ 
	\frac{ {\rm d} n^{}_{\mu\mu} }{{\rm d} z} &=& \sum_{I=1}^{3} \left\{ \varepsilon^{\left(I\right)}_{\mu\mu} D^{}_I \left( n^{}_{N^{}_I} - n^{\rm eq}_{N^{}_I} \right)
	- W^{}_I \left[ \left|C^{0}_{I\mu}\right|^2 n^{}_{\mu\mu} + {\rm Re} \left( C^{0}_{I\mu} C^{0}_{Ie} n^{}_{\mu e} +  C^{0}_{I\mu} C^{0}_{I\tau} n^{*}_{\tau \mu} \right) \right]
	\right\}  \;,
\nonumber	\\ 
	\frac{ {\rm d} n^{}_{\tau\tau} }{{\rm d} z} &=& \sum_{I=1}^{3} \left\{ \varepsilon^{\left(I\right)}_{\tau\tau} D^{}_I \left( n^{}_{N^{}_I} - n^{\rm eq}_{N^{}_I} \right)
	- W^{}_I \left[ \left|C^{0}_{I\tau}\right|^2 n^{}_{\tau\tau} + {\rm Re} \left( C^{0}_{I\tau} C^{0}_{Ie} n^{}_{\tau e} +  C^{0}_{I\tau} C^{0}_{I\mu} n^{}_{\tau \mu} \right) \right]
	\right\}  \;,
\nonumber	\\ 
	\frac{ {\rm d} n^{}_{\tau\mu} }{{\rm d} z} &=& \sum_{I=1}^{3} \bigg\{ \varepsilon^{\left(I\right)}_{\tau\mu} D^{}_I \left( n^{}_{N^{}_I} - n^{\rm eq}_{N^{}_I} \right)
	- \frac{1}{2} W^{}_I \bigg[ n^{}_{\tau\mu} \left( \left|C^{0}_{I\tau}\right|^2 + \left|C^{0}_{I\mu}\right|^2 \right)
	+  C^{0*}_{I\mu} C^{0}_{I\tau} \left( n^{}_{\tau\tau} + n^{}_{\mu\mu} \right) \nonumber \\
	&&\hspace{1cm} + C^{0*}_{Ie} C^{0}_{I\tau} n^{*}_{\mu e} +  C^{0*}_{I\mu} C^{0}_{Ie} n^{}_{\tau e} \bigg]
	\bigg\} - \left[ \frac{ {\rm Im} \left( \Lambda^{}_\tau \right) }{ Hz}  + \frac{ {\rm Im} \left( \Lambda^{}_\mu \right) }{ Hz} \right] n^{}_{\tau\mu} \;,
\nonumber	\\ 
	\frac{ {\rm d} n^{}_{\tau e} }{{\rm d} z} &=& \sum_{I=1}^{3} \bigg\{ \varepsilon^{\left(I\right)}_{\tau e} D^{}_I \left( n^{}_{N^{}_I} - n^{\rm eq}_{N^{}_I} \right)
	- \frac{1}{2} W^{}_I \bigg[ n^{}_{\tau e} \left( \left|C^{0}_{Ie }\right|^2 + \left|C^{0}_{I\tau}\right|^2 \right)
	+  C^{0*}_{Ie} C^{0}_{I\tau} \left( n^{}_{ee} + n^{}_{\tau\tau} \right) \nonumber \\
	&&\hspace{1cm} + C^{0*}_{I\mu} C^{0}_{I\tau} n^{}_{\mu e} +  C^{0*}_{Ie} C^{0}_{I\mu} n^{}_{\tau \mu} \bigg]
	\bigg\} - \frac{ {\rm Im} \left( \Lambda^{}_\tau \right) }{ Hz} n^{}_{\tau e} \;,
\nonumber	\\ 
	\frac{ {\rm d} n^{}_{\mu e} }{{\rm d} z} &=& \sum_{I=1}^{3} \bigg\{ \varepsilon^{\left(I\right)}_{\mu e} D^{}_I \left( n^{}_{N^{}_I} - n^{\rm eq}_{N^{}_I} \right)
	- \frac{1}{2} W^{}_I \bigg[ n^{}_{\mu e} \left( \left|C^{0}_{Ie }\right|^2 + \left|C^{0}_{I\mu}\right|^2 \right)
	+  C^{0*}_{Ie} C^{0}_{I\mu} \left( n^{}_{ee} + n^{}_{\mu\mu} \right) \nonumber \\
	&&\hspace{1cm} + C^{0*}_{Ie} C^{0}_{I\tau} n^{*}_{\tau \mu} +  C^{0*}_{I\tau} C^{0}_{I\mu} n^{}_{\tau e} \bigg]
	\bigg\} - \frac{ {\rm Im} \left( \Lambda^{}_\mu \right) }{ Hz} n^{}_{\mu e} \;,
	\label{2.7}
\end{eqnarray}
where $C^{0}_{I\alpha}$ have been defined in Eq.~(\ref{2.6}).
In these equations, $n^{\rm eq}_{N^{}_I}$ are the equilibrium number density of three right-handed neutrinos. The decay (washout) parameters $ D^{}_I\left( W^{}_I \right)$  are given by
\begin{eqnarray}
	D^{}_I \left( z\right) = K^{}_I x^{}_I z \frac{ \mathcal{K}^{}_1 \left( z^{}_I \right) }{ \mathcal{K}^{}_2 \left( z^{}_I \right) }  \;,\qquad
	W^{}_I \left( z\right) = \frac{1}{4} K^{}_I \sqrt{x^{}_I} \; \mathcal{K}^{}_1 \left( z^{}_I \right) z^3_I \;,
	\label{2.8}
\end{eqnarray}
with $\mathcal{K}^{}_1$ and $\mathcal{K}^{}_2$ being the modified Bessel functions of the second kind and
\begin{eqnarray}
	x^{}_{I} \equiv \frac{M^2_I}{M^2_1} \;,\qquad  z^{}_I \equiv z \sqrt{x^{}_I} \;,\qquad
	K^{}_I \equiv \frac{\Gamma^{}_I}{H \left(T=M^{}_1 \right)} \;.
	\label{2.9}
\end{eqnarray}
Moreover, the Hubble expansion rate $H$ is given by
\begin{eqnarray}
	H = \sqrt{ \frac{4\pi^3g^{}_*}{45} } \frac{M^2_1}{z^2M^{}_{\rm Pl}} \;,
	\label{2.10}
\end{eqnarray}
in which $M^{}_{\rm Pl} = 1.22 \times 10^{19}$GeV is the Planck mass and $g^{}_* \simeq 106.75$ or 228.75 (in the SM and MSSM frameworks, respectively) is the number of relativistic degrees of freedom in the early Universe. Last but not least,
$ \Lambda^{}_\alpha$ is the self-energy of $\alpha$-flavored leptons, and its imaginary part is given by
\begin{eqnarray}
	{\rm Im} \left( \Lambda^{}_\alpha \right) \simeq 8 \times 10^{-3} y^2_\alpha T \;,
	\label{2.11}
\end{eqnarray}
with $y^{}_\alpha$ being the charged-lepton Yukawa coupling.

\subsection{Some formulas about RGE effects}

As mentioned in the introduction section, the flavor-symmetry models which contain three degenerate right-handed neutrinos serve as a unique basis for realizations of the tri-resonant leptogenesis scenario after the right-handed neutrino masses receive certain corrections so that their degeneracy is lifted to an appropriate extent. And the RGE effects between the flavor-symmetry and leptogenesis scales can play the role of generating the desired right-handed neutrino mass splitings. In the literature, most of the flavor-symmetry models have been formulated in the MSSM framework for the following reason: in order to break the flavor symmetry in a proper way, one needs to introduce some flavon fields which transform as multiplets of the flavor symmetry and develop particular VEV alignments; and the most
popular and perhaps natural approach to derive the desired flavon VEV alignments is
provided by the so-called F-term alignment mechanism which is realized in a supersymmetric setup \cite{FS}. In particular, the modular flavor-symmetry models have been formulated in the MSSM framework since the modular flavor symmetry needs supersymmetry to preserve the holomorphicity of the
modular form \cite{modular}. For these reasons, we will perform our study in the MSSM framework.

In the MSSM framework, the relevant renormalization-group equations have been given in Ref.~\cite{Casas:1999ac}.
First of all, the lepton fields can be transformed by a unitary transformation to the basis in which the right-handed neutrino mass matrix is diagonal. In this basis, the RGE evolution of three right-handed neutrino masses $M^{}_{I}$ and the neutrino Yukawa coupling matrix $Y$ can be obtained by using the methods outlined in \cite{rgeL,Babu:1987im}
\begin{eqnarray}
	&&	4 \pi^2 \frac{ {\rm d} M^{}_{I} }{ {\rm d} t} = \left( Y Y^{\dagger}_{} \right)^{}_{II} M^{}_{I} \;, \nonumber \\
	&&	16 \pi^2 \frac{ {\rm d} Y }{ {\rm d} t} = - T^* Y - Y \left[ \alpha -  \left( 3 Y^\dagger Y + Y^\dagger_l Y^{}_l \right) \right] \;,
	\label{2.2.1}
\end{eqnarray}
where $t = \ln \left(M^{}_0/ \mu\right)$ has been defined, with $M^{}_0$ and $\mu$ denoting the right-handed neutrino mass scale and renormalization scale, respectively. And one has
$\alpha = 3 g^2_2 + \frac{3}{5}g^2_1 - {\rm Tr} \left( 3 Y^\dagger_{u} Y^{}_{u} + Y^\dagger Y \right)$, where $g^{}_2$ and $g^{}_1$ are the $\rm SU\left(2\right)^{}_{\rm L}$ and $\rm U\left(1\right)^{}_{\rm Y}$ gauge coupling constants,
and $Y^{}_{u, l}$ are the Yukawa coupling matrices for the up-type quarks and charged leptons.
$T$ is a $3\times 3$ anti-Hermitian matrix and its non-zero entries are given by
\begin{eqnarray}
T^{}_{IJ} = - T^{*}_{JI} = \frac{ M^{}_{J} + M^{}_{I} }{M^{}_{J} - M^{}_{I}} {\rm Re} [ ( Y  Y^\dagger )^{}_{IJ} ] +
{\rm i} \frac{ M^{}_{J} - M^{}_{I} }{M^{}_{J} + M^{}_{I}} {\rm Im} [ ( Y  Y^\dagger )^{}_{IJ} ] \;,\qquad {\rm for} \;\; I\neq J \;.
	\label{2.2.2}
\end{eqnarray}
It is immediate to see that $T^{}_{IJ}$ is singular at the scale where one has $M^{}_{J} = M^{}_{I}$, unless $ {\rm Re} [ ( Y  Y^\dagger )^{}_{IJ} ] = 0$ holds.
Fortunately, due to the exact mass degeneracy of three right-handed neutrinos, one has a freedom of rotation $N \to O N$ with $O^T O =I$ for them,
which does not affect the unity right-handed neutrino mass matrix, but rotates the Yukawa coupling matrix to the following form
\begin{eqnarray}
Y \to Y^{\prime} = O Y \;.
	\label{2.2.3}
\end{eqnarray}
Such a freedom can be utilized to assure that Eq.~(\ref{2.2.1}) is non-singular \cite{Turzynski:2004xy}, namely,
\begin{eqnarray}
{\rm Re} \left[ Y^{\prime} Y^{\prime\dagger} \right] = {\rm Re} \left[ O^\dagger Y Y^\dagger O \right] = 0 \;.
	\label{2.2.4}
\end{eqnarray}

\section{Study for a model realizing the TM1 mixing}

In this section, we perform the study for a neutrino flavor-symmetry model realizing the popular TM1 mixing \cite{TM1}. To be specific, we will consider the concrete model presented in section~3 of Ref.~\cite{Li:2013jya}, which is constructed based on the ${\rm S}^{}_4$ flavor symmetry in the MSSM framework. ${\rm S}^{}_4$ is the permutation group of 4 objects with 24 elements, and it has two singlet representations ${\bf 1}$ and ${\bf 1^\prime}$, one two-dimensional representation ${\bf 2}$ and two three-dimensional representations ${\bf 3}$ and ${\bf 3^\prime}$. In this model both the three left-handed lepton doublets and three right-handed neutrinos constitute a ${\bf 3^\prime}$ representation of ${\rm S}^{}_4$, while three right-handed charged leptons are trivially ${\bf 1}$ representations of ${\rm S}^{}_4$. In order to be able to yield certain predictions for the CP phases at the same time, the flavor symmetry is extended in such a way that a generalized CP symmetry \cite{gcp} is also included. As is commonly done in the paradigm of flavor symmetries, the full symmetries including both the flavor and CP symmetries are broken (by the flavon VEVs) into different subgroups in the neutrino and charged-lepton sectors, and the mismatch between them gives rise to certain predictions for the lepton mixing angles and CP phases.

In the model under consideration, thanks to the residual flavor and CP symmetries, the charged-lepton mass matrix is constrained to be diagonal, and the Dirac neutrino mass matrix is constrained into the following form
\begin{eqnarray}
	&& M^{}_{\rm D} = \frac{y^{}_1 v^{}_u v^{}_\xi}{\Lambda^{}_{\rm FS}} \left[
	\begin{pmatrix}
		1 & 0 & 0  \\ 0 & 0 & 1 \\ 0 & 1 & 0
	\end{pmatrix}
	+ x \begin{pmatrix}
		0 & 1 & 1  \\ 1 & 1 & 0 \\ 1 & 0 & 1
	\end{pmatrix}
	+ {\rm i} y \begin{pmatrix}
		0 & 1 & -1  \\ -1 & 0 & -2 \\ 1 & 2 & 0
	\end{pmatrix}
	\right] \;,
	\label{3.1}
\end{eqnarray}
while the right-handed neutrino mass matrix takes the same form as in Eq.~(\ref{6}).
Here $v^{}_{u}$ and $v^{}_{\xi}$ are the VEVs of the up-type Higgs and flavon fields, $y^{}_1$, $x$ and $y$ are real parameters. Given the above forms of $M^{}_{\rm D}$ and $M^{}_{\rm R}$, the seesaw formula yields a light neutrino mass matrix as
\begin{eqnarray}
	M^{}_{\nu} = a \left( \begin{matrix}
		2 & -1 & -1 \cr
		-1 & 2 & -1 \cr
		-1 & -1 & 2
	\end{matrix} \right) +
	b \left( \begin{matrix}
		1 & 0 & 0 \cr
		0 & 0 & 1 \cr
		0 & 1 & 0
	\end{matrix} \right) +
	c \left( \begin{matrix}
		0 & 1 & 1 \cr
		1 & 1 & 0 \cr
		1 & 0 & 1
	\end{matrix} \right) +
	{\rm i} d \left( \begin{matrix}
		0 & 1 & -1 \cr
		1 & 2 & 0 \cr
		-1 & 0 & -2
	\end{matrix} \right) \;,
	\label{3.2}
\end{eqnarray}
with
\begin{eqnarray}
	&& a= m^{}_0 y^2 \;, \hspace{1cm} b= - m^{}_0 (1+ 2 x^2 + 4 y^2) \;, \nonumber \\
	&& c= - m^{}_0 (2 x + x^2 + y^2) \;, \hspace{1cm} d = - 3 m^{}_0 x y \;,
	\label{3.3}
\end{eqnarray}
and $m^{}_0 = y^2_1 v^2_u v^2_\xi/(M^{}_0 \Lambda^2_{\rm FS})$. It is easy to verify that such an $M^{}_\nu$ not only keeps invariant under the TM1 transformation but also under the $\mu$-$\tau$ reflection transformation \cite{MTR}:
\begin{eqnarray}
	&& M^{}_\nu = R^{T}_{\rm TM1} M^{}_\nu R^{}_{\rm TM1}
	\hspace{0.5cm} {\rm with} \hspace{0.5cm}
	R^{}_{\rm TM1}= - \frac{1}{3} \left( \begin{array}{ccc}
		1 & 2 & 2 \cr
		2 & -2 & 1 \cr
		2 & 1 & -2 \cr
	\end{array} \right)  \;, \nonumber \\
	&& M^{}_\nu = R^{T}_{\mu\tau} M^{*}_\nu R^{}_{\mu\tau}
	\hspace{0.5cm} {\rm with} \hspace{0.5cm}
	R^{}_{\mu\tau}= \left( \begin{array}{ccc}
		1 & 0 & 0 \cr
		0 & 0 & 1 \cr
		0 & 1 & 0 \cr
	\end{array} \right)  \;.
	\label{3.4}
\end{eqnarray}
Correspondingly, the resulting neutrino mixing not only has the features of the TM1 mixing
but also the features of the $\mu$-$\tau$ reflection symmetry:
\begin{eqnarray}
	\theta^{}_{23} = \frac{\pi}{4} \;, \hspace{1cm} \delta = \pm \frac{\pi}{2} \;,
	\hspace{1cm} \alpha^{}_{21} = 0 \ {\rm or} \ \frac{\pi}{2} \;, \hspace{1cm} \alpha^{}_{31} = 0 \ {\rm or} \ \frac{\pi}{2} \;.
	\label{3.5}
\end{eqnarray}
Given that the considered model is subject to further constraints on top of the TM1 mixing, if it can successfully reproduce the observed value of $Y^{}_{\rm B}$ with the help of the RGE effects, then the models that are only subject to the TM1 mixing necessarily can also do so.

It is found that this model can only be phenomenologically viable in the NO case and the best-fit values of $x$ and $y$ are $x=-0.050$ and $y=\pm 0.233$ (corresponding to $\delta = \pm \pi/2$) \cite{Li:2013jya}. And the parameter combination $y^{}_1 v^{}_u v^{}_\xi /\Lambda^{}_{\rm FS}$ in $M^{}_{\rm D}$ can be expressed as $\sqrt{m^{}_{ee} M^{}_0/(1+2x^2+2y^2)}$ with $m^{}_{ee} \simeq 57.9$ meV (i.e., the effective neutrino mass for neutrinoless double beta decays).
The most stringent bounds on neutrinoless double beta decay lifetimes come from the KamLAND-Zen~\cite{KamLAND-Zen:2024eml} and GERDA~\cite{GERDA:2020xhi} experiments, which constrain the effective Majorana mass as $m^{}_{ee}<$  (28-122)meV and $m^{}_{ee}<$  (79-180) meV, respectively.
The model also predicts the masses of the three neutrinos: $m^{}_1 = 57.284$ meV, $m^{}_2 = 57.930$ meV and $m^{}_3 = 75.488$ meV. For comparison, the latest Planck results constrain the sum of neutrino masses to $\sum^{}_i m^{}_i<$ (0.12-0.60)eV~\cite{Planck:2018vyg}.
Furthermore, the Yukawa coupling matrix can be obtained as $Y = M^{}_{\rm D}/v^{}_u$ with $v^{}_u = v \tan \beta/\sqrt{1+\tan^2\beta}$. Here $\tan \beta$ is the ratio of the VEV of the up-type Higgs field $H^{}_u$ to that of the down-type Higgs field $H^{}_d$.
Note that in the MSSM a moderately large $\tan \beta$ value (e.g., 30) is favored, for which the relation $v^{}_u \simeq v$ holds.

\begin{figure}[htbp]
	\centering
	\includegraphics[width=6.0in]{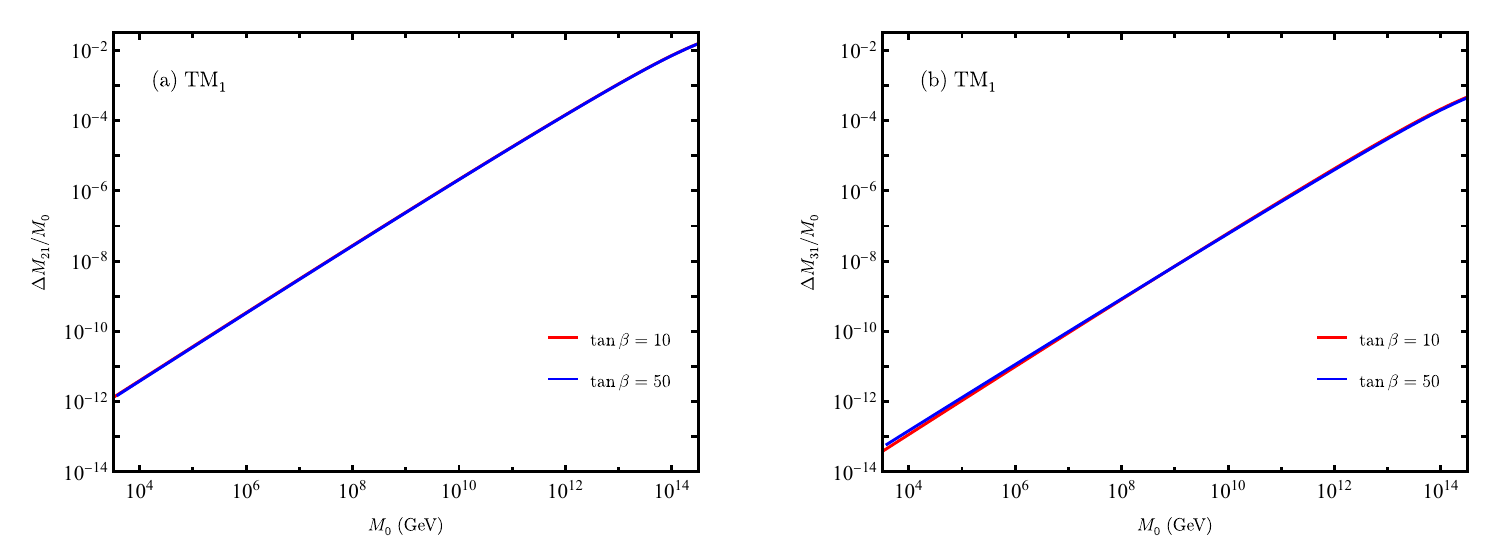}
	\caption{ For the TM1 model in section~3, the mass splittings among the three right-handed neutrinos generated from the RGE effects, as functions of $M^{}_0$ for the benchmark values of $\tan \beta =10$ and 50.}
	\label{fig1}
\end{figure}

Now, we are ready to study the RGE induced tri-resonant leptogenesis for the above model: at the flavor-symmetry scale $\Lambda^{}_{\rm FS}$, three right-handed neutrinos are completely degenerate, prohibiting leptogenesis to proceed; if there is a considerable gap between $\Lambda^{}_{\rm FS}$ and the leptogenesis scale (roughly $M^{}_0$), then the RGE effects from the former scale to the latter scale will lift the degeneracy among three right-handed neutrino masses, inducing a tri-resonant leptogenesis to work. Given that most parameters of the model have been fixed as described above, there are only three free parameters relevant to our study: $M^{}_0$, $\Lambda^{}_{\rm FS}$ and $\tan \beta $. Considering that the RGE effects only depend on $\Lambda^{}_{\rm FS}$ in a logarithmic manner (and we have numerically checked that the dependence of the final results on the concrete value of $\Lambda^{}_{\rm FS}$ is weak),
in the following numerical calculations we will take $\Lambda^{}_{\rm FS}=10^{16}$ GeV (i.e., around the grand unification scale) as a benchmark value.
Now that $M^{}_0$ and $\tan \beta $ are the remaining free parameters (note that the value of $\tan \beta $ will directly affect the sizes of the charged-lepton Yukawa coupling and consequently the flavor effects), in Figure~\ref{fig1} we have shown the RGE-induced mass splittings of the three right-handed neutrinos as functions of $M^{}_0$ for the benchmark values of $\tan \beta =10$ and 50. The results show that the ratios $\Delta M^{}_{IJ} / M^{}_0$ (for $\Delta M^{}_{IJ} \equiv M^{}_I - M^{}_J$) are nearly linearly proportional to $M^{}_0$ and  insensitive to $\tan \beta $. This is because $\Delta M^{}_{IJ} / M^{}_0$ are proportional to $YY^\dagger$ (see Eq.~(\ref{2.2.1})) which in turn is proportional to $M^{}_0$. And $YY^\dagger$ is almost independent of $\tan \beta $ when it takes values from 10 to 50 (as mentioned in the end of last paragraph).
	
\begin{figure}[htbp]
	\centering
	\includegraphics[width=6.0in]{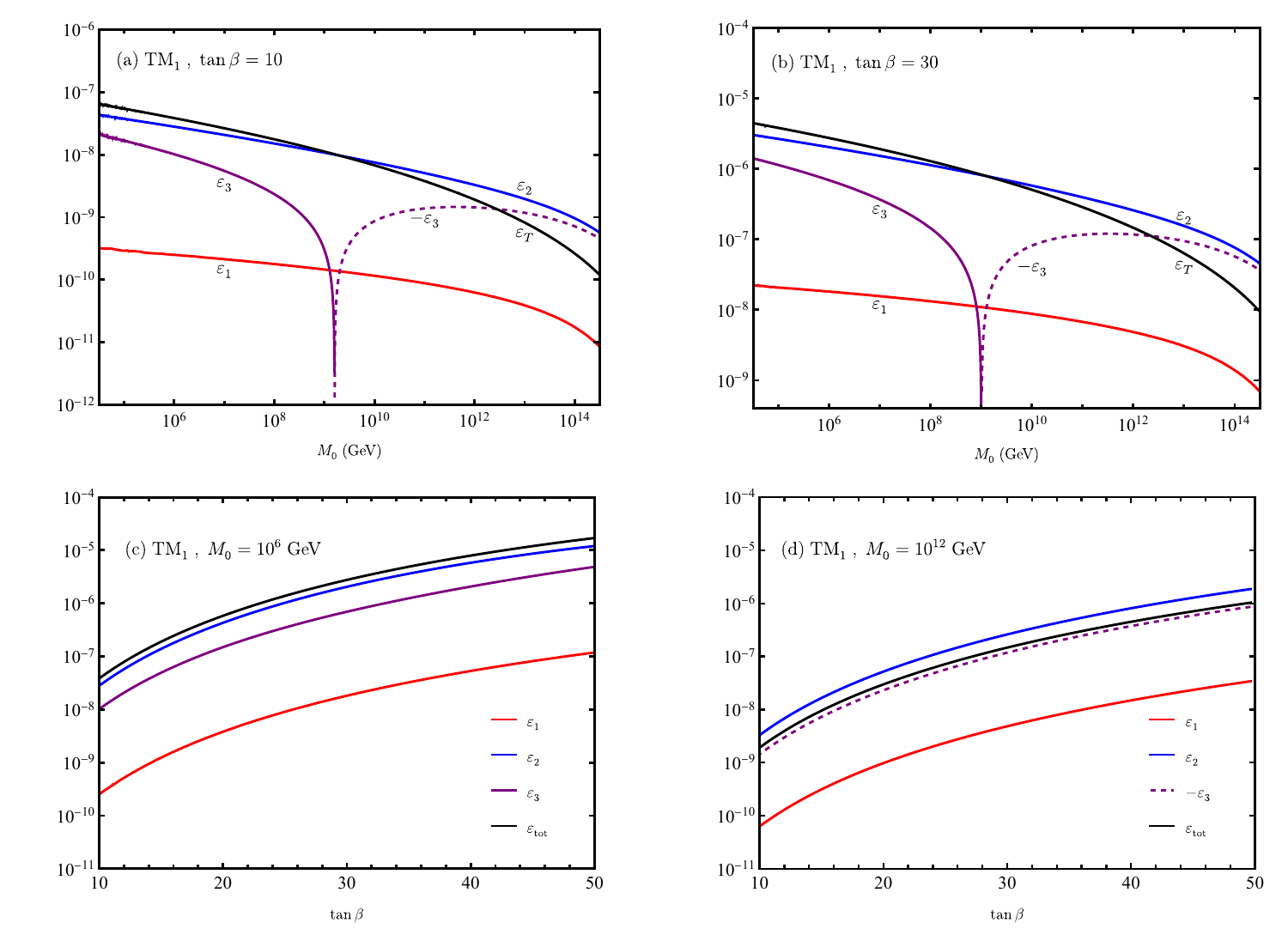}
	\caption{For the TM1 model in section~3, the resulting CP asymmetries for the decays of $N^{}_{I}$ as functions of $M^{}_0$ and $\tan \beta$.}
	\label{fig2}
\end{figure}

\begin{figure}[htbp]
	\centering
	\includegraphics[width=3.0in]{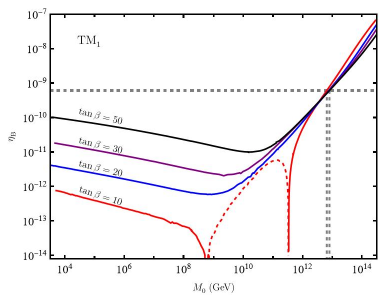}
	\caption{For the TM1 model in section~3, the resulting $\eta^{}_{\rm B}$ as functions of $M^{}_0$ for some benchmark values of $\tan \beta $. The horizontal gray dashed line shows the observed value of $\eta^{}_{\rm B}$. }
	\label{fig3}
\end{figure}

Figure~\ref{fig2} has shown the resulting CP asymmetries $\varepsilon^{}_{I} = {\rm Tr} \; \varepsilon^{\left(I\right)}_{\alpha \beta}$ for the decays of $N^{}_{I}$ as functions of $M^{}_0$ and $\tan \beta$.
The results show that $\varepsilon^{}_{I}$ and $\varepsilon^{}_{T} = \sum_{I} \varepsilon^{}_{I}$ decrease with the increasing of $M^{}_0$, and increase with $\tan \beta$.
And the sign of $\varepsilon^{}_{3}$ changes from positive to negative at $M^{}_0 \simeq 10^{12}$ GeV. Please keep in mind that here and in the following, for negative results, we have used the dashed lines to show their absolute values.
Finally, Figure~\ref{fig3} has shown the resulting $\eta^{}_{\rm B}$ as functions of $M^{}_0$ for the benchmark values of $\tan \beta= 10$, 20, 30 and 50.
One can see that for $M^{}_0 > 10^{12}$ GeV the four curves nearly converge into a single one.
This means that in such a mass range the resulting $\eta^{}_B$ is almost independent of $\tan \beta$.
And the observed value of $\eta^{}_{\rm B}$ can be successfully reproduced for $M^{}_0 \sim 7 \times 10^{12}$ GeV.

\section{Study for modular ${\rm A}^{}_4$ symmetry model}

In Ref.~\cite{Ding:2019zxk}, the authors have presented a comprehensive analysis of neutrino masses and mixing in theories with modular ${\rm A}^{}_4$ symmetry. ${\rm A}^{}_4$ is the symmetry group of the even permutations of 4 objects, and it has three singlet representations ${\bf 1}$, ${\bf 1^\prime}$, ${\bf 1^{\prime \prime}}$ and one three-dimensional representation ${\bf 3}$. Note that ${\rm A}^{}_4$ is the smallest non-Abelian finite group which admits a three-dimensional representation (a desired feature for describing the flavor mixing of three generation fermions). In this model, both the three left-handed lepton doublets and three right handed neutrinos constitute the ${\bf 3}$ representation of ${\rm A}^{}_4$. In the case that the weights of the lepton doublets and right-handed neutrinos are $(k^{}_L, k^{}_{N^c_{}}) = (2, 0)$, the modular invariance constrains the Dirac neutrino mass matrix to the following form
\begin{eqnarray}
M^{}_D =  v^{}_u \begin{pmatrix}
		2 g^{}_1 Y^{}_1 & \left( -g^{}_1+g^{}_2 \right) Y^{}_3 &  \left( -g^{}_1-g^{}_2 \right) Y^{}_2 \\
	\left( -g^{}_1-g^{}_2 \right) Y^{}_3	 & 2 g^{}_1 Y^{}_2  & \left( -g^{}_1+g^{}_2 \right) Y^{}_1 \\	\left( -g^{}_1+g^{}_2 \right) Y^{}_2 & \left( -g^{}_1-g^{}_2 \right) Y^{}_1  & 2 g^{}_1 Y^{}_3
	\end{pmatrix}   \;,
\label{4.1}
\end{eqnarray}
while the right-handed neutrino mass matrix takes the same form as in Eq.~(\ref{6}).
On the other hand, in the case that three right-handed charged leptons are successively assigned to the singlet representations ${\bf 1}$, ${\bf 1^{\prime\prime}}$ and ${\bf 1^\prime}$ of ${\rm A}^{}_4$, the charged-lepton mass matrix takes a form as
\begin{eqnarray}
M^{}_{l} = v^{}_d \begin{pmatrix}
	\alpha Y^{}_1 & \alpha Y^{}_3	 & \alpha Y^{}_2 	  \\
	\beta Y^{}_2   & 	\beta Y^{}_1 & \beta Y^{}_3	 \\	
	\gamma Y^{}_3  & 	\gamma Y^{}_2  & 	\gamma Y^{}_1
\end{pmatrix} .
\label{4.2}
\end{eqnarray}
In the above two equations, $v^{}_u$ and $v^{}_d$ are respectively the VEVs of the up-type and down-type Higgs fields, $g^{}_{1}$, $g^{}_{2}$, $\alpha$, $\beta$ and $\gamma$ are simply dimensionless parameters, while $Y^{}_i$ are modular forms of level 3 and they can be expressed in terms of the Dedekind eta-function
\begin{eqnarray}
\eta(\tau) \equiv q^{1/24} \prod_{n=1}^{\infty} \left(1 - q^n \right) \hspace{0.5cm} {\rm with} \hspace{0.5cm} q = e^{ {\rm i} 2\pi \tau} \;,
\label{4.3}
\end{eqnarray}
and its derivative (with ${\rm d}\eta/{\rm d}\tau$ being denoted as $\eta^\prime$) as
\begin{eqnarray}
	&& Y^{}_1(\tau) = \frac{\rm i}{2 \pi} \left[\frac{\eta^{\prime}(\tau / 3)}{\eta(\tau / 3)}+\frac{\eta^{\prime}(\tau/3+1/3)}{\eta(\tau/3+1/3)}+\frac{\eta^{\prime}(\tau/3+2/3)}{\eta( \tau/3+2/ 3)}-\frac{27 \eta^{\prime}(3 \tau)}{\eta(3 \tau)}\right] \;, \\
	&& Y^{}_2(\tau)=\frac{-i}{\pi}\left[\frac{\eta^{\prime}(\tau / 3)}{\eta(\tau / 3)}+\omega^2 \frac{\eta^{\prime}( \tau/3+1/3)}{\eta( \tau/3+1/3)}+\omega \frac{\eta^{\prime}( \tau/3+2/ 3)}{\eta( \tau/3 +2/ 3)}\right]  \;, \\
	&& Y^{}_3(\tau)=\frac{-i}{\pi}\left[\frac{\eta^{\prime}(\tau / 3)}{\eta(\tau / 3)}+\omega \frac{\eta^{\prime}(\tau/3 +1 / 3)}{\eta( \tau/3+1/ 3)}+\omega^2 \frac{\eta^{\prime}( \tau/3 + 2/ 3)}{\eta( \tau/3 +2/ 3)}\right]  \;,
\label{4.4}
\end{eqnarray}
with $\omega = e^{ {\rm i} 2 \pi / 3 }$. One can see that this model totally contains 8 real parameters: ${\rm Re}(\tau)$, ${\rm Im}(\tau)$, $\beta/\alpha$, $\gamma/\alpha$, $|g^{}_2/g^{}_1|$, ${\rm arg} (g^{}_2/g^{}_1)$, $\alpha v^{}_d$ and $g^2_1 v^2_u /M^{}_0$. The three parameters $\alpha v_d$, $\beta/\alpha$ and $\gamma/\alpha$ can be fixed by fitting with the values of three charged lepton masses. The remaining parameters serve to describe the three neutrino masses, three neutrino mixing angles and three CP phases present in the neutrino mixing matrix.
The detailed numerical calculation shows that this model can be consistent with the neutrino oscillation experimental results in both the NO and IO cases. And the resulting predictions for the best-fit values and the allowed ranges of the input parameters and observables are collected in Table~\ref{a4model} here.

\begin{table}[htpb]
\centering
\renewcommand{\arraystretch}{0.8}
\resizebox{1\textwidth}{!}{\begin{tabular}{|c|cc|cc|} \hline
&  \multicolumn{2}{c|}{NO}                &  \multicolumn{2}{c|}{IO} \\ \cline{2-5}
& Best-fit  &   Allowed regions & Best-fit  &  Allowed regions \\  \hline

${\rm Re}(\tau)$ & 0.0129 & $[0, 0.431]$  & 0.096 & $[0, 0.102]$  \\
${\rm Im}(\tau)$ & 1.824 &  $[0.91, 1.16]\cup[1.31,1.86] $ & 0.987 &  $[0.98, 1.049]\cup[1.052,1.109]$   \\
$\beta/\alpha$ & 205.720 & $[192.39,215]\cup[3054.25, 4093.49]$ & 79.472 & $[59.68,86.37]\cup[892.67, 1446.02]$  \\
$\gamma/\alpha$  & 3612.07 & $[192.4,215]\cup[3066.47, 4092.98]$  & 1232.57 & $[60.97,86.34]\cup[870.16, 1443.84]$ \\
$|g^{}_2/g^{}_1|$   & 2.410 & $[2.398,2.71]\cup[2.95,3.86]$  & 2.093 & $[1.038, 2.453]$ \\
$\arg{(g^{}_2/g^{}_1)}$ & 6.267 & $[0, 0.49] \cup [6.23,2\pi]$ & 4.715 & $[1.33, 1.83]\cup[4.29,5]$ \\
$\alpha v_d$/MeV  & 0.5179 & ---   & 1.167 & ---  \\
$(g^2_1 v_u^2/M^{}_0)$/eV   & 0.0111 & ---  & 0.004 & ---   \\ \hline
$m^{}_e/m^{}_{\mu}$   & 0.0048 & $[0.0046, 0.0050]$ & 0.0048 & $[0.0046, 0.0050]$  \\
$m^{}_{\mu}/m^{}_{\tau}$  & 0.0564 & $[0.0520, 0.0610]$  & 0.0565 &$[0.0520, 0.0610]$   \\ \hline
$\sin^2\theta^{}_{12}$  & 0.3096 & $[0.2750, 0.3500]$  & 0.3100 & $[0.2750, 0.3500]$   \\
$\sin^2\theta^{}_{13}$  & 0.0226  & $[0.02045, 0.02439]$ & 0.02264 & $[0.02068, 0.02463]$ \\
$\sin^2 \theta^{}_{23}$ & 0.4638 & $[0.4180, 0.4676]$ & 0.584 & $[0.423, 0.629]$  \\
$\delta/\pi$    & 1.486 &  $[0.19,0.77]\cup [1.23,1.84]$  & 1.458 &  $[0.068, 1.933]$  \\
$\alpha^{}_{21}/\pi$   & 0.068 & $[0,0.23]\cup [1.77,2]$& 0.138 & $[0,0.19]\cup[1.8,2]$   \\
$\alpha^{}_{31}/\pi$  & 0.948 & $[0.683, 1.311]$  & 0.997 & $[0, 2]$  \\ \hline
$m^{}_1$/eV   & 0.0430 & $[0.0225, 0.0478]$  & 0.0494 & $[0.0464, 0.0526]$  \\
$m^{}_2$/eV   & 0.0439 & $[0.0241, 0.0485]$ & 0.0501 & $[0.0472, 0.0533]$  \\
$m^{}_3$/eV   & 0.0661 & $[0.0524, 0.0716]$  & 0.0013 & $[0.0007, 0.0015]$  \\
$\sum^{}_i m^{}_i$/eV &  0.153  &  $[0.0991, 0.1679]$  &   0.1008  &  $[0.0942, 0.1074]$   \\
$ m^{}_{ee} $/eV   & 0.0436  &   $[0.0202, 0.0483]$   &   0.0475  &   $[0.0439, 0.0516]$   \\
\hline

$\chi^2_{\mathrm{min}}$                        & 30.72 &  ---               & $10^{-7}$ &  ---      \\ \hline
\end{tabular} }
\caption{For the modular ${\rm A}^{}_4$ symmetry model in section~4, the predictions for the best-fit values and the allowed ranges of the input parameters and observables in the NO and IO cases. These results are taken from Tables~6 and 10 in Ref.~\cite{Ding:2019zxk}.}
\label{a4model}
\end{table}

\begin{figure}[htbp]
	\centering
	\includegraphics[width=6.0in]{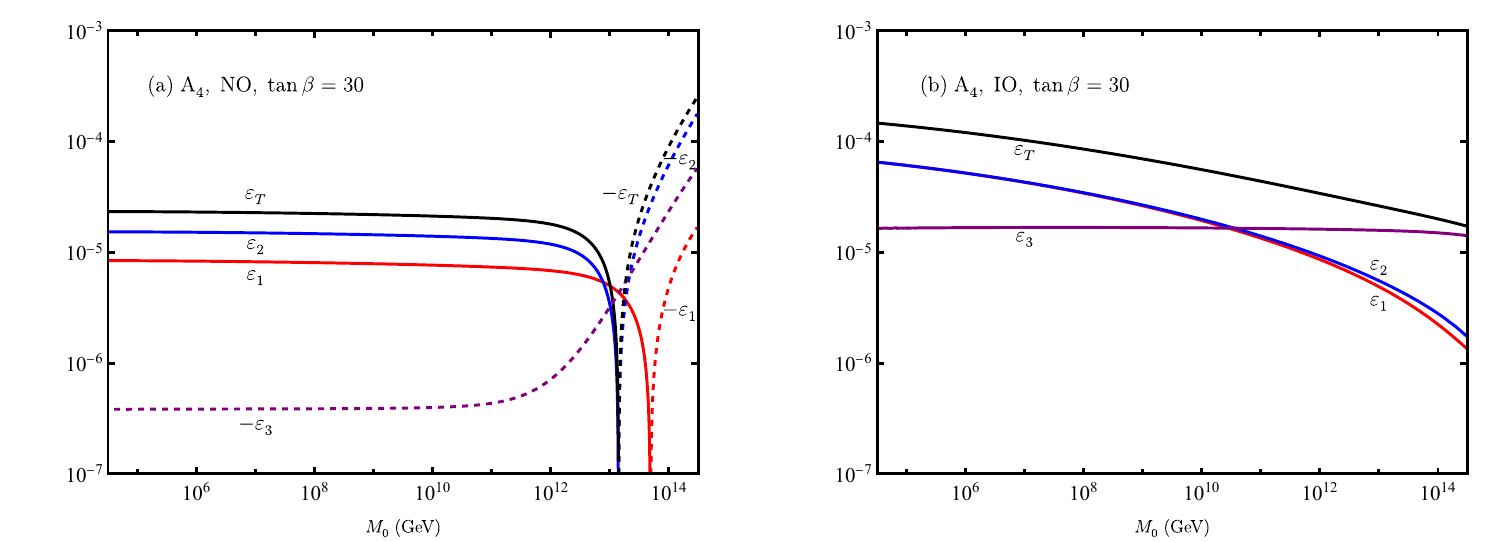}
	\caption{For the modular ${\rm A}^{}_4$ symmetry model in section~4, the resulting CP asymmetries for the decays of $N^{}_{I}$ as functions of $M^{}_0$ for the benchmark value of $\tan \beta = 30$ in the NO (left panel) and IO (right panel) cases.}
	\label{fig5}
\end{figure}

\begin{figure}[htbp]
	\centering
	\includegraphics[width=6.0in]{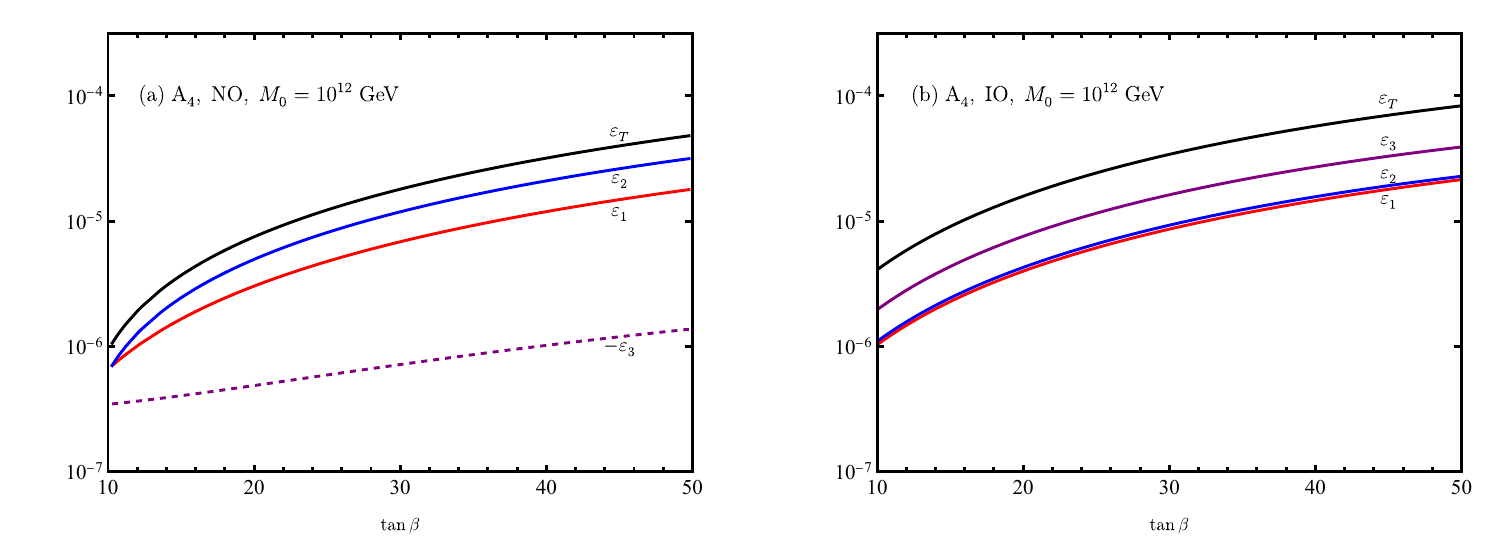}
	\caption{For the modular ${\rm A}^{}_4$ symmetry model in section~4, the resulting CP asymmetries for the decays of $N^{}_{I}$ as functions of $\tan \beta$ for the benchmark value of $ M^{}_0= 10^{12}$ GeV in the NO (left panel) and IO (right panel) cases.}
	\label{fig6}
\end{figure}

\begin{figure}[htbp]
	\centering
	\includegraphics[width=6.0in]{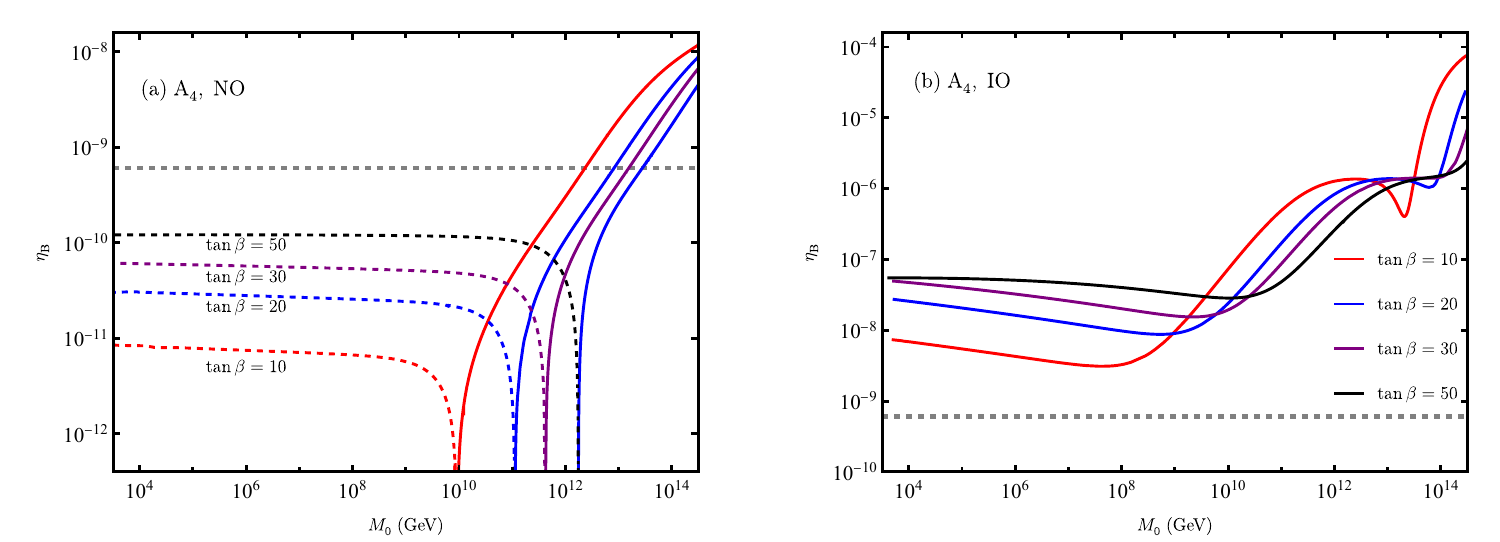}
	\caption{For the modular ${\rm A}^{}_4$ symmetry model in section~4, the resulting $\eta^{}_{\rm B}$ as functions of $M^{}_0$ for the benchmark values of $\tan \beta =$10, 20, 30 and 50 in the NO (left panel) and IO (right panel) cases. The horizontal gray dashed line shows the observed value of $\eta^{}_{\rm B}$.}
	\label{fig7}
\end{figure}

\begin{figure}[htbp]
	\centering
	\includegraphics[width=3.0in]{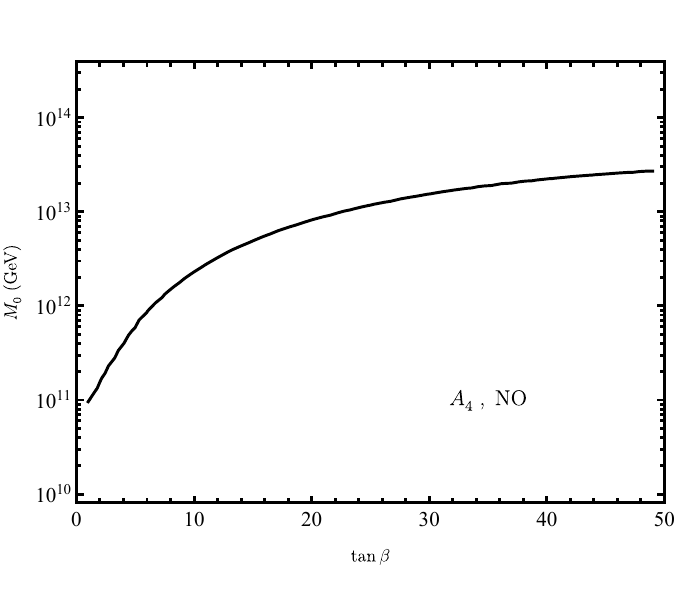}
	\caption{For the modular ${\rm A}^{}_4$ symmetry model in section~4, in the NO case, the values of $M^{}_0$ versus $\tan \beta $ that allow for a successful reproduction of the observed value of $\eta^{}_{\rm B}$.}
	\label{fig8}
\end{figure}

Now, we study the RGE induced tri-resonant leptogenesis for the above model.
For the purpose of illustration, Figure~\ref{fig5} has shown the CP asymmetries for the decays of $N^{}_{I}$ as functions of $M^{}_0$ for the  benchmark value of $\tan \beta = 30$.
In the NO case, for $M^{}_0 < 10^{12}$ GeV, the CP asymmetries are nearly independent of $M^{}_0$.
But for $M^{}_0 > 10^{12}$ GeV, the absolute value of $\varepsilon^{}_3$ increases with $M^{}_0$, while $\varepsilon^{}_1$, $\varepsilon^{}_2$ and $\varepsilon^{}_T$ also do so after a sign reversal from being positive to negative. In the IO case, the CP asymmetries keep their signs unchanged and change very slowly as $M^{}_0$ increases. Figure~\ref{fig6} has further shown the CP asymmetries for the decays of $N^{}_{I}$ as functions of $\tan \beta$ for the benchmark value of $ M^{}_0= 10^{12}$ GeV. In both the NO and IO cases, the CP asymmetries increase with $\tan \beta$.
Figure~\ref{fig7} has shown the resulting $\eta^{}_{\rm B}$ as functions of $M^{}_0$ for some benchmark values of $\tan \beta $.
As one can see, in the NO case, the observed value of $\eta^{}_{\rm B}$ can be successfully reproduced when $M^{}_0$ is approximately between $10^{12}$ GeV and  $10^{13}$ GeV.
Figure~\ref{fig8} has further shown the values of $M^{}_0$ versus $\tan \beta $ that allow for a successful reproduction of the observed value of $\eta^{}_{\rm B}$.
But in the IO case, the resulting $\eta^{}_{\rm B}$ is always larger than the observed value, since the corresponding CP asymmetries are too large (see the right panel of Figure~\ref{fig5}).

\section{Study for modular ${\rm S}^{}_4$ symmetry model}

In Ref.~\cite{s4}, the authors have presented a comprehensive analysis of neutrino masses and mixing in theories with modular ${\rm S}^{}_4$ symmetry. In the case that the three right-handed neutrinos have a weight $k^{}_{N^c} =0$ and constitute the ${\bf 3}$ or ${\bf 3^\prime}$ representation of ${\rm S}^{}_4$, the right-handed neutrino mass matrix will take a same form as in Eq.~(\ref{6}). When the three left-handed lepton doublets have a weight $k^{}_{L} =2$ and constitute the same three-dimensional
representation of ${\rm S}^{}_4$ as the three right-handed neutrinos, the Dirac neutrino mass matrix will take a form as
\begin{eqnarray}
M^{}_{\rm D} = g v^{}_u \left[
\begin{pmatrix}
0 & Y^{}_1 & Y^{}_2 \\
Y^{}_1 & Y^{}_2 & 0 \\
Y^{}_2 & 0 & Y^{}_1
\end{pmatrix}
+ \frac{g^\prime}{g} \begin{pmatrix}
0 & Y^{}_5 & -Y^{}_4 \\
- Y^{}_5 & 0 & Y^{}_3 \\
Y^{}_4 & -Y^{}_3 & 0
\end{pmatrix}
\right] \;.
\label{5.1}
\end{eqnarray}
If the three left-handed lepton doublets constitute a three-dimensional representation of ${\rm S}^{}_4$ different from the three right-handed neutrinos, then the Dirac neutrino mass matrix will take a form as
\begin{eqnarray}
M^{}_{\rm D} = g v^{}_u \left[
\begin{pmatrix}
0 & -Y^{}_1 & Y^{}_2 \\
-Y^{}_1 & Y^{}_2 & 0 \\
Y^{}_2 & 0 & -Y^{}_1
\end{pmatrix}
+ \frac{g^\prime}{g} \begin{pmatrix}
2Y^{}_3 & -Y^{}_5 & -Y^{}_4 \\
-Y^{}_5 & 2Y^{}_4 & -Y^{}_3 \\
-Y^{}_4 & -Y^{}_3 & 2Y^{}_5
\end{pmatrix}
\right] \;.
\label{5.2}
\end{eqnarray}
On the other hand, for the minimal (in terms of weights) possibility, the charged-lepton mass matrix takes a form as
\begin{eqnarray}
M^{}_l = v^{}_d
\begin{pmatrix}
\alpha Y^{}_3 & \alpha Y^{}_5 & \alpha Y^{}_4 \\
\beta \left(Y^{}_1 Y^{}_4 - Y^{}_2 Y^{}_5\right) & \beta \left(Y^{}_1 Y^{}_3 - Y^{}_2 Y^{}_4\right) & \beta \left(Y^{}_1 Y^{}_5 - Y^{}_2 Y^{}_3\right) \\
\gamma \left(Y^{}_1 Y^{}_4 + Y^{}_2 Y^{}_5\right) & \gamma \left(Y^{}_1 Y^{}_3 + Y^{}_2 Y^{}_4\right) & \gamma \left(Y^{}_1 Y^{}_5 + Y^{}_2 Y^{}_3\right) \\
\end{pmatrix} \;.
\label{5.3}
\end{eqnarray}
In the above three equations, $g$, $g^\prime$, $\alpha$, $\beta$ and $\gamma$ are simply dimensionless parameters, while $Y^{}_i$ are modular forms of weight 2 and level 4 whose explicit expressions have been formulated in Ref.~\cite{s4modular}.

One can see that this model also contains 8 real parameters in total: $\alpha v_d$, $\beta/\alpha$,
$\gamma/\alpha$, $g^2 v^2_u /M^{}_0$, $|g^\prime/g|$, ${\rm arg}(g^\prime/g)$, ${\rm Re}(\tau)$ and ${\rm Im}(\tau)$. The numerical calculation shows that this model can be consistent with the neutrino oscillation experimental results. For $M^{}_{\rm D}$ in Eq.~(\ref{5.1}), the model can be consistent with the neutrino oscillation experimental results only in the NO case, and
the numerical search finds two pairs of distinct local minima of $\Delta \chi^2$ corresponding to two pairs of distinct values of $\tau$. The two minima in each pair lead to opposite values of the Dirac and Majorana phases, but the same values of all other observables. The cases belonging to the first pair are denoted as Case A, while those belonging to the second pair are denoted as Case B. For these two cases, the resulting predictions for the best-fit values and the allowed ranges of the input parameters and observables are collected in Table~\ref{s4model1} here \cite{s4}. For $M^{}_{\rm D}$ in Eq.~(\ref{5.2}), the model can be consistent with the neutrino oscillation experimental results in both the NO and IO cases. In the IO case, the numerical search finds two pairs of distinct local minima of $\Delta \chi^2$ corresponding to two pairs of distinct values of $\tau$, and the cases belonging to these two pairs are respectively denoted as Case C and Case D. In the NO case, the numerical search only finds one pair of local minima of $\Delta \chi^2$ corresponding to one pair of distinct values of $\tau$, and the cases belonging to this pair are denoted as Case E. For these three cases, the resulting predictions for the best-fit values and the allowed ranges of the input parameters and observables are collected in Table~\ref{s4model2} here \cite{s4}.

\begin{table}[htpb]
	\centering
	\renewcommand{\arraystretch}{1}
	\resizebox{1\textwidth}{!}{\begin{tabular}{|c|cc|cc|} \hline
	&  \multicolumn{2}{c|}{Case A (NO)}                &  \multicolumn{2}{c|}{Case B (NO)} \\ \cline{2-5}
	& Best-fit  &   Allowed regions & Best-fit  &  Allowed regions \\  \hline
	${\rm Re}(\tau)$ & $\pm 0.1045$ &  $\pm (0.09378 - 0.1128)$ & $\mp 0.109$ & $\mp (0.103 - 0.1197)$ \\
	${\rm Im}(\tau)$ & 1.01  & $1.004 - 1.018$ & 1.005 & $0.9988 - 1.008$ \\
	$\beta/\alpha$ & 9.465  & $7.693 - 12.39$ & 0.03306  & $0.02529 - 0.04074$ \\
	$\gamma/\alpha$ & 0.002205  & $0.001941 - 0.002472$ & 0.0001307 & $0.0000982 - 0.0001663$ \\
	${\rm Re}(g'/g) $ & 0.233  & $-0.02544 - 0.4417$ & 0.4097  & $0.3241 - 0.5989$ \\
	${\rm Im}(g'/g) $ & $\pm 0.4924$  & $\pm (-0.6046 - 0.5751)$ & $\mp 0.5745$ & $\mp (0.5436 - 0.5944)$ \\
	$ \alpha v_d$ [MeV] & 53.19  & --- & 893.2 & --- \\
	$ g^2 v_u^2 / M^{}_0$ [eV] & 0.00467 & --- & 0.004014 & --- \\	
	\hline
	$m_e/m_{\mu}$ & 0.004802  & $0.00422 - 0.005383$ & 0.004802 &  $0.004211 - 0.005384$ \\
	$m_{\mu} / m_{\tau}$ & 0.0565  & $0.04317 - 0.06961$ & 0.05649  & $0.04318 - 0.06962$ \\
	$\sin^2 \theta_{12}$ & 0.305  & $0.2656 - 0.3449$  & 0.305 & $0.2662 - 0.3455$ \\
	$\sin^2 \theta_{13}$ & 0.02125 & $0.01912 - 0.02383$ & 0.02125 & $0.01914 - 0.02383$ \\
	$\sin^2 \theta_{23}$ & 0.551  & $0.4838 - 0.5999$ & 0.551 & $0.4322 - 0.601$ \\
	\hline
	$m_1$ [eV] & 0.01746  & $0.01185 - 0.02143$ & 0.02074 & $0.01918 - 0.02428$ \\
	$m_2$ [eV] & 0.01945  & $0.01473 - 0.02307$ & 0.02244 & $0.02101 - 0.02574$ \\
	$m_3$ [eV] & 0.05288  & $0.05075 - 0.05452$ & 0.05406 & $0.05314 - 0.05577$ \\
  $\sum^{}_i m^{}_i$ [eV] &   0.0898  &   $[0.07735, 0.09887]$   &  0.09724   &  $[0.0935, 0.1056]$    \\
 $m^{}_{ee}$ [eV]    &   0.01699  &   $[0.01177, 0.02002]$   &  0.01983  &  $[0.01847, 0.02275]$  \\
	$\delta/\pi$ & $\pm 1.314$  & $\pm (1.249 - 1.961)$ & $\pm 1.919$ & $\pm (1.882 - 1.977)$ \\
	$\alpha^{}_{21}/\pi$ & $\pm 0.302$  & $\pm (0.2748 - 0.3708)$ & $\pm 1.704$ & $\pm (1.681 - 1.722)$ \\
	$\alpha^{}_{31}/\pi$ & $\pm 0.8716$ & $\pm (0.7973 - 1.635)$ & $\pm 1.539$ & $\pm (1.484 - 1.618)$ \\
	\hline
	\(N\sigma \equiv \sqrt{\Delta \chi^2}\) & 0.02005 & --- & 0.02435 & --- \\
	\hline
	\end{tabular} }
	\caption{ For the modular ${\rm S}^{}_4$ symmetry model in section~5 with $M^{}_{\rm D}$ in Eq.~(\ref{5.1}), in the phenomenologically viable NO case, the predictions for the best-fit values and the allowed ranges of the input parameters and observables in Case A and Case B. These results are taken from Tables~5a-5b in Ref.~\cite{s4}. }
	\label{s4model1}
\end{table}

\begin{table}[htpb]
	\centering
	\renewcommand{\arraystretch}{1.2}
	\resizebox{1\textwidth}{!}{\begin{tabular}{|c|cc|cc|cc|} \hline
	&  \multicolumn{2}{c|}{Case C (IO)}                &  \multicolumn{2}{c|}{Case D (IO)}  &  \multicolumn{2}{c|}{Case E  (NO) } \\ \cline{2-7}
	& Best-fit  &   Allowed regions & Best-fit  &  Allowed regions & Best-fit & Allowed regions \\  \hline
	${\rm Re}(\tau)$ & $\mp 0.1435$ &  $\mp (0.1222 - 0.168)$ & $\pm 0.179$ & $\pm (0.1589 - 0.199)$ & $\mp 0.4996$ & $\mp (0.48 - 0.5084)$ \\
	${\rm Im}(\tau)$ & 1.523  & $1.088 - 1.594$ & 1.397 &  $1.236 - 1.529$ & 1.309 & $1.246 - 1.385$ \\
	$\beta/\alpha$ & 17.82  & $9.32 - 23.66$ & 15.35 & $10.79 - 21.09$ & 0.000243 & $0.0002004 - 0.0002864$ \\
	$\gamma/\alpha$ & 0.003243  & $0.00227 - 0.003733$ & 0.002924 & $0.002443 - 0.003459$ & 0.03335 & $0.02799 - 0.03926$ \\
	${\rm Re}(g'/g) $ &  $-0.8714$  & $-(0.7956 - 1.148)$ & $-1.32$ & $-(1.131 - 1.447)$ & $-0.06454$ & $-(0.01697 - 0.1215)$ \\
	${\rm Im}(g'/g) $ & $\mp 2.094$ & $\mp (1.409 - 2.182)$ & $\pm 1.733$ & $\pm (1.306 - 2.017)$ & $\mp 0.569$ & $\mp (0.4572 - 0.6564)$ \\
	$\alpha v_d$ [MeV] &  71.26  & --- & 68.42 & --- & 1125 & ---\\
	$g^2 v_u^2/ M^{}_0$ [eV] & 0.004087 & --- & 0.00447 & --- & 0.0087 & ---\\	
	\hline
	$m_e/m_{\mu}$ & 0.004797  & $0.004215 - 0.005378$ & 0.004786 & $0.004221 - 0.005386$ & 0.004797 & $0.004393 - 0.005197$ \\
	$m_{\mu} / m_{\tau}$ & 0.05655  & $0.04348 - 0.0698$ & 0.0554 & $0.04343 - 0.06968$ & 0.05626 & $0.04741 - 0.0654$ \\
	$\sin^2 \theta_{12}$ & 0.303 & $0.2657 - 0.3436$ & 0.3031  & $0.2657 - 0.3436$ & 0.311 & $0.2895 - 0.3375$ \\
	$\sin^2 \theta_{13}$ & 0.02175 & $0.01957 - 0.0242$ & 0.02184 & $0.01954 - 0.0242$ & 0.02185 & $0.02041 - 0.02351$ \\
	$\sin^2 \theta_{23}$ & 0.5571 & $0.4551 - 0.6026$ & 0.5577 & $0.5482 - 0.6013$ & 0.4469 & $0.43 - 0.4614$ \\
	\hline
	$m_1$ [eV] & 0.0513  & $0.04882 - 0.05207$ & 0.05122 & $0.05023 - 0.05212$ & 0.01774 & $0.01703 - 0.01837$ \\
	$m_2$ [eV] & 0.05201 & $0.04958 - 0.05274$ & 0.05193 & $0.05098 - 0.05279$ & 0.0197 & $0.01906 - 0.02025$ \\
	$m_3$ [eV] & 0.01512 & $0.00316 - 0.0163$ & 0.01495 & $0.01223 - 0.01649$ & 0.05299 & $0.05251 - 0.05346$ \\
	 $\sum^{}_i m^{}_i$ [eV] &  0.1184 & $0.102-0.1208$ &  0.1181 & $0.1139 -0.1212$ &  0.09043 &  $0.08874-0.09195$ \\
	 $ m^{}_{ee} $ [eV] &  0.0263 & $0.02288-0.04551$ & 0.0310 &  $0.02515-0.03677$ &  0.006967 &  $0.006482-0.007288$ \\
	$\delta/\pi$ & $\pm 1.098$ & $\pm (0.98 - 1.289)$ & $\pm 1.384$ & $\pm (1.271 - 1.437)$ & $\pm 1.601$ & $\pm (1.287 - 1.828)$ \\
	$\alpha_{21}/\pi$ & $\pm 1.241$ & $\pm (1.113 - 1.758)$ & $\pm 1.343$ & $\pm (1.171 - 1.479)$ & $\pm 1.093$ & $\pm (0.8593 - 1.178)$ \\
	$\alpha_{31}/\pi$ & $\pm 0.2487$ & $\pm (0.069 - 0.346)$ & $\pm 0.806$ & $\pm (0.448 - 1.149)$ & $\pm 0.7363$ & $\pm (0.3334 - 0.9643)$ \\
	\hline
	\(N\sigma \equiv \sqrt{\Delta \chi^2}\) & 0.0357 & --- & 0.3811 & --- & 2.147 & --- \\
	\hline
	\end{tabular} }
	\caption{ For the modular ${\rm S}^{}_4$ symmetry model in section~5 with $M^{}_{\rm D}$ in Eq.~(\ref{5.2}), in their respective phenomenologically viable neutrino mass ordering cases, the predictions for the best-fit values and the allowed ranges of the input parameters and observables in Case C, Case D and Case E. These results are taken from Tables~5c-5e in Ref.~\cite{s4}.  }
	\label{s4model2}
\end{table}

\begin{figure}[htbp]
	\centering
	\includegraphics[width=6.0in]{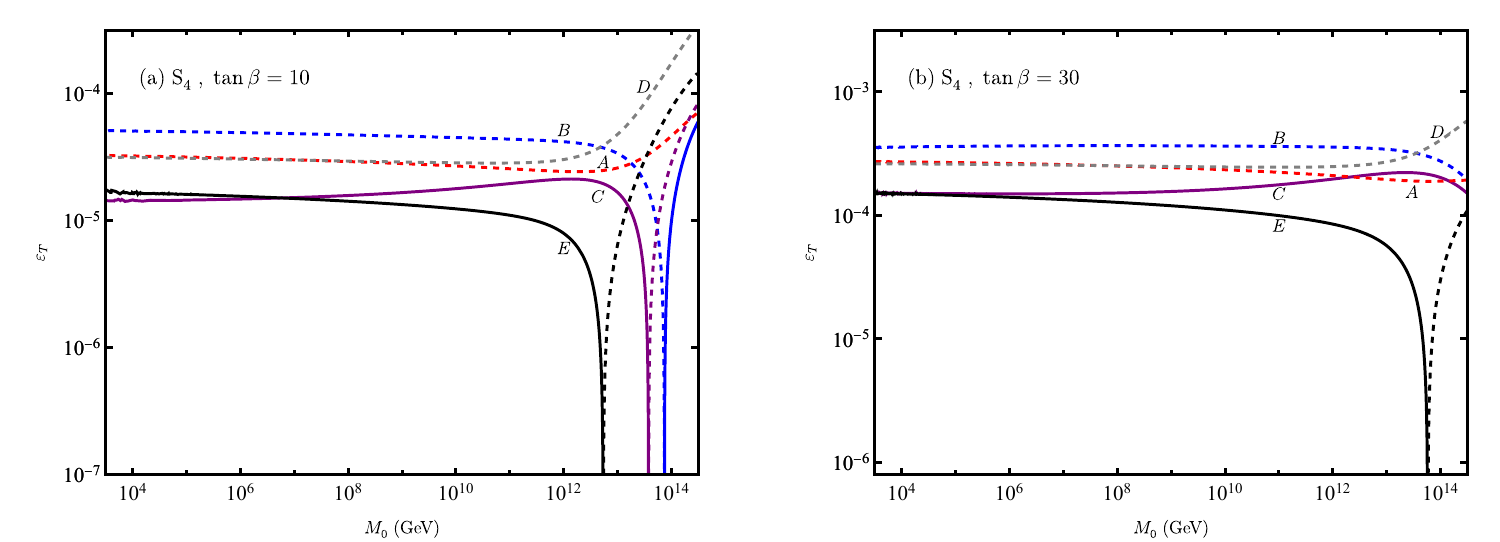}
	\caption{ For the five cases of modular ${\rm S}^{}_4$ symmetry model in section~5, the total CP asymmetries $\varepsilon^{}_{T}$ for the decays of $N^{}_I$ as functions of $M^{}_0 $ for the benchmark values of $\tan \beta = 10$ (left panel) and $30$ (right panel).}
	\label{fig9}
\end{figure}	

\begin{figure}[htbp]
	\centering
	\includegraphics[width=6.0in]{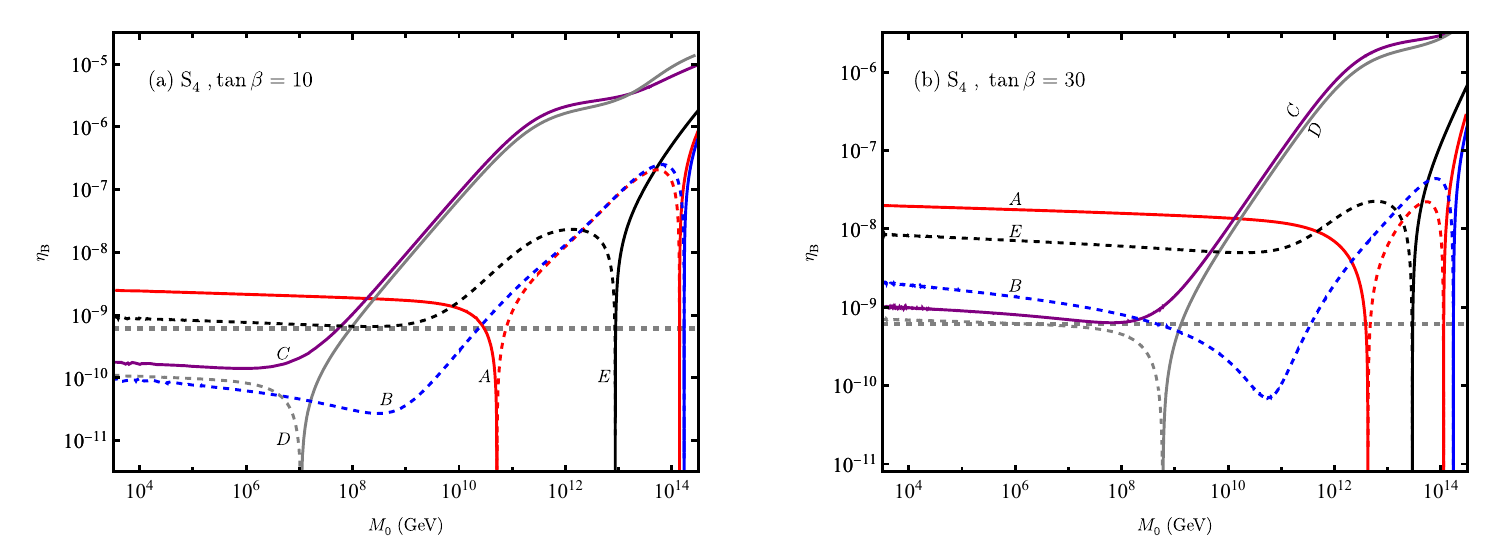}
	\caption{ For the five cases of modular ${\rm S}^{}_4$ symmetry model in section~5, the resulting $\eta^{}_{\rm B}$ as functions of $M^{}_0$ for the benchmark values of $\tan \beta = 10$ (left panel) and $30$ (right panel). The horizontal gray dashed line shows the observed value of $\eta^{}_{\rm B}$.}
	\label{fig10}
\end{figure}	

\begin{figure}[htbp]
	\centering
	\includegraphics[width=3.0in]{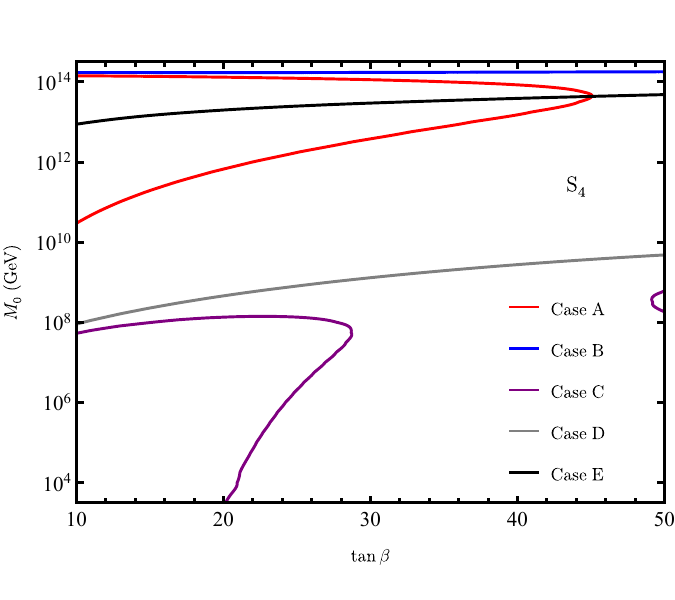}
	\caption{ For the five cases of modular ${\rm S}^{}_4$ symmetry model in section~5, the values of $M^{}_0$ versus $\tan \beta $ that allow for a successful reproduction of the observed value of $\eta^{}_{\rm B}$.}
	\label{fig11}
\end{figure}	

Now, we study the RGE induced tri-resonant leptogenesis for these five cases.
Figure~\ref{fig9} has shown the total CP asymmetries $\varepsilon^{}_{T}$ for the decays of $N^{}_I$ as functions of $M^{}_0$ for the benchmark values of $\tan \beta = 10$ and 30.
For $M^{}_0 < 10^{12}$ GeV, all of the CP asymmetries are almost independent of $M^{}_0$, while they change very sharply for $M^{}_0 > 10^{12}$ GeV.
Figure~\ref{fig10} has shown the resulting $\eta^{}_{\rm B}$ as functions of $M^{}_0$ also for the benchmark values of $\tan \beta = 10$ and 30.
On can see that the observed value of $\eta^{}_{\rm B}$ has chance to be reproduced successfully for appropriate values of $M^{}_0$ and $\tan \beta$. This can be seen more clearly from Figure~\ref{fig11}
which has shown the values of $M^{}_0$ versus $\tan \beta $ that allow for a successful reproduction of the observed value of $\eta^{}_{\rm B}$.
The results show that for Case A and Case C there may exist two allowed values of $M^{}_0$ for each allowed value of $\tan \beta$. The results show that for Case A and Case C there is an upper bound on $\tan \beta$ (about 45 and 28, respectively) and for each allowed value of $\tan \beta$ there may exist two allowed values of $M^{}_0$.
Importantly, Case C allows relatively low values of $M^{}_0$, which may be reachable for future direct measurements. For the other cases, the allowed values of $M^{}_0$ (about $10^{14}$, $10^9$ and $10^{13}$ GeV for Case B, Case D and Case E, respectively) are almost independent of $\tan \beta$.

\section{Study for modular ${\rm A}^{}_5$ symmetry model}

In Ref.~\cite{a5}, the authors have presented a comprehensive analysis of neutrino masses and mixing in theories with modular ${\rm A}^{}_5$ symmetry. ${\rm A}^{}_5$ is the symmetry group of even permutations of 5 objects, and it has one singlet representation ${\bf 1}$, two three-dimensional representations ${\bf 3}$ and ${\bf 3}^\prime$, one four-dimensional representation ${\bf 4}$ and one five-dimensional representation ${\bf 5}$. In that paper, the authors have constructed modular-invariant models in which all leptons are collected into ${\bf 3}$ or ${\bf 3}^\prime$ multiplets of ${\rm A}^{}_5$, containing the three generations of each type of field. The neutrino sector has been taken to be minimal and it only depends on the modulus $\tau$ and an overall scale. But modular forms do not appear in the charged-lepton sector, which instead contains two extra flavons. In the case that the three right-handed neutrinos have a weight $k^{}_{N^c} =0$ and constitute the ${\bf 3}$ representation of ${\rm A}^{}_5$, the right-handed neutrino mass matrix will take a same form as in Eq.~(\ref{6}). When the three left-handed lepton doublets have a weight $k^{}_{L} =-2$ and constitute ${\bf 3}^\prime$ representation of ${\rm A}^{}_5$, the Dirac neutrino mass matrix will take a form as
\begin{eqnarray}
M^{}_{\rm D} = y^{}_0 v^{}_u \left(\begin{array}{ccc}
        \sqrt{3} Y^{}_1 & Y^{}_5 & Y^{}_2 \\
        Y^{}_4 & -\sqrt{2} Y^{}_3 & -\sqrt{2} Y^{}_5 \\
        Y^{}_3 & -\sqrt{2} Y^{}_2 & -\sqrt{2} Y^{}_4
      \end{array}\right) \;.
\label{6.1}
\end{eqnarray}
On the other hand, for the minimal (in terms of weights) possibility, the charged-lepton mass matrix takes a form as
\begin{equation}
M^{}_l = v^{}_d
  \left(
    \begin{array}{ccc}
      \alpha + 4 \gamma (1 - \varphi^{}_2 \varphi^{}_3) &
      (\beta + 6 \gamma) \varphi^{}_3 &
      (-\beta + 6 \gamma) \varphi^{}_2 \\
      (-\beta + 6 \gamma) \varphi^{}_3 &
      6 \gamma \varphi_3^2 &
      \alpha + \beta - 2 \gamma (1 - \varphi^{}_2 \varphi^{}_3) \\
      (\beta + 6 \gamma) \varphi^{}_2 &
      \alpha - \beta - 2 \gamma (1 - \varphi^{}_2 \varphi^{}_3) &
      6 \gamma \varphi_2^2
    \end{array}
  \right)~~~.
\label{6.2}
\end{equation}
In the above two equations, $y^{}_0$, $\alpha$, $\beta$ and $\gamma$ are simply dimensionless parameters, $\varphi^{}_2$ and $\varphi^{}_3$ are the VEVs of relevant flavon fields and will be treated as free parameters, while $Y^{}_i$ are modular forms of weight 2 and level 5 whose explicit expressions have been formulated in Ref.~\cite{a5modular}.

Similar to the models in sections~4 and 5, in the present model the charged-lepton masses are essentially controlled by $\alpha v^{}_d$, $\beta/\alpha$ and $\gamma/\alpha$, while the neutrino masses and mixing angles are mainly governed by $y^2_0 v^2_u/M^{}_0$, $\tau$, $\varphi^{}_2$ and $\varphi^{}_3$.
The number of parameters is further reduced by enforcing the CP conservation which restricts all the Yukawa couplings together with $\varphi^{}_2$ and $\varphi^{}_3$ to be real (with the VEV of $\tau$ serving as the only source of CP violation). The numerical calculation shows that this model can be consistent with the neutrino oscillation experimental results only in the NO case and it further predicts $m^{}_1 =0$. The resulting predictions for the best-fit values of the input parameters and observables are collected in Table~\ref{a5model} here.

\begin{table}[h!]
\footnotesize
\begin{center}
\noindent\makebox[\textwidth]{
\begin{tabular}{|c|c|c|c|c|c|c|c|}
\hline
Re($\tau$) & Im$(\tau)$ & $\varphi_2$  & $\varphi_3$  & $\alpha \cos \beta \cdot 10^{3}$ & $\beta/\alpha$ & $\gamma/\alpha$ & $y^2_0 v^2 \sin^2 \beta$/$M^{}_0$ (eV$^{-1}$)
\rule[-2ex]{0pt}{5ex}\\
\hline
 $-$0.3615 & 0.2412 & 0.04759  & 0.3731 & 3.368 & 1.310 & $-$0.2413 & 0.0001639
\rule[-2ex]{0pt}{5ex}\\
\hline
\end{tabular}}
\vspace{0.5cm}

\noindent\makebox[\textwidth]{
	\begin{tabular}{|c|c|c|c|c|c|c|c|c|c|}
		\hline
$m_1 ({\rm meV}^{})$& $m_2 ({\rm meV}^{})$ & $m_3 ({\rm meV}^{})$ &
 $m^{}_{ee} ({\rm meV}^{})$ &  $\sin ^{2} \theta_{12}$ &$\sin ^{2} \theta_{13}$ &$\sin ^{2} \theta_{23}$ &$\delta / \pi$ & $(\alpha_{21}$-$\alpha_{31})/\pi$ & $\chi ^{2} _{\textrm{min}}$
		\rule[-2ex]{0pt}{5ex}\\
		\hline
		0 & 8.60 & 50.2 & 1.3  & 0.292  & 0.0228  & 0.449 &  1.63  &  1.68 & 11.1
		\rule[-2ex]{0pt}{5ex}\\
		\hline
\end{tabular}}

\caption{For the modular ${\rm A}^{}_5$ symmetry model in section~6, in the phenomenologically viable NO case, the predictions for the best-fit values of the input parameters and observables. These results are taken from Table~8 in Ref.~\cite{a5}.}
\label{a5model}
\end{center}
\end{table}

\begin{figure}[htbp]
	\centering
	\includegraphics[width=6.0in]{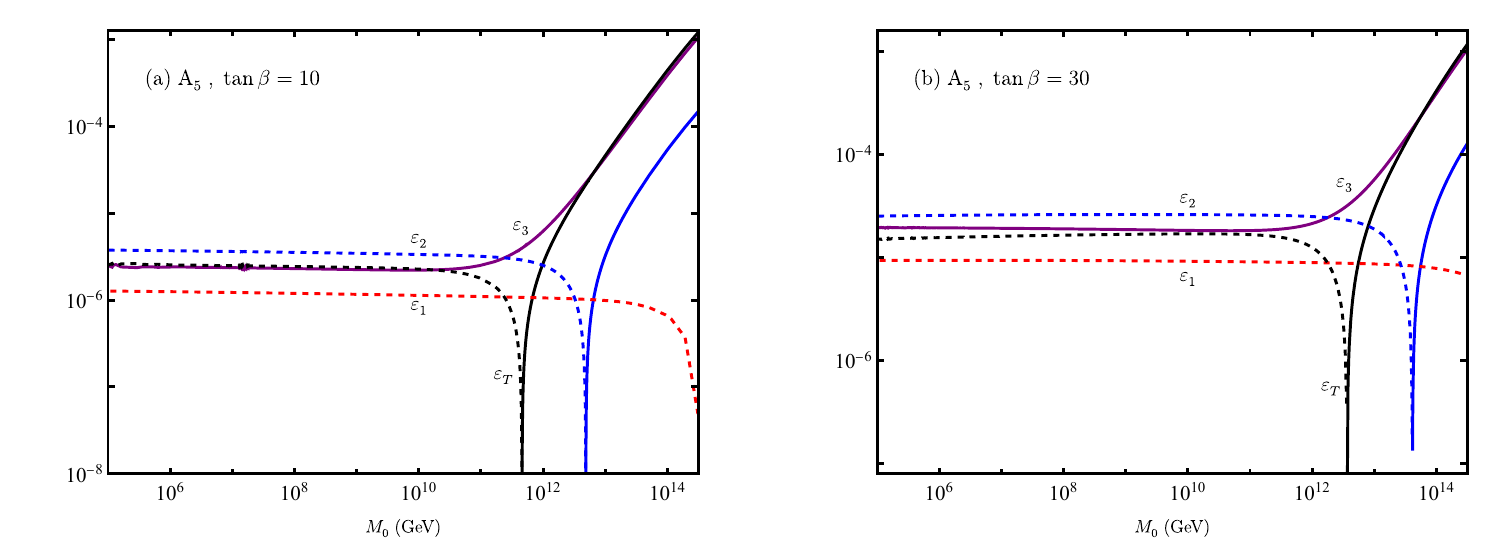}
	\caption{For the modular ${\rm A}^{}_5$	symmetry model in section~6, the resulting CP asymmetries for the decays of $N^{}_{I}$ as functions of $M^{}_0$ for the benchmark values of $\tan \beta =$10 (left panel) and 30 (right panel).}
	\label{fig12}
\end{figure}

\begin{figure}[htbp]
	\centering
	\includegraphics[width=6.0in]{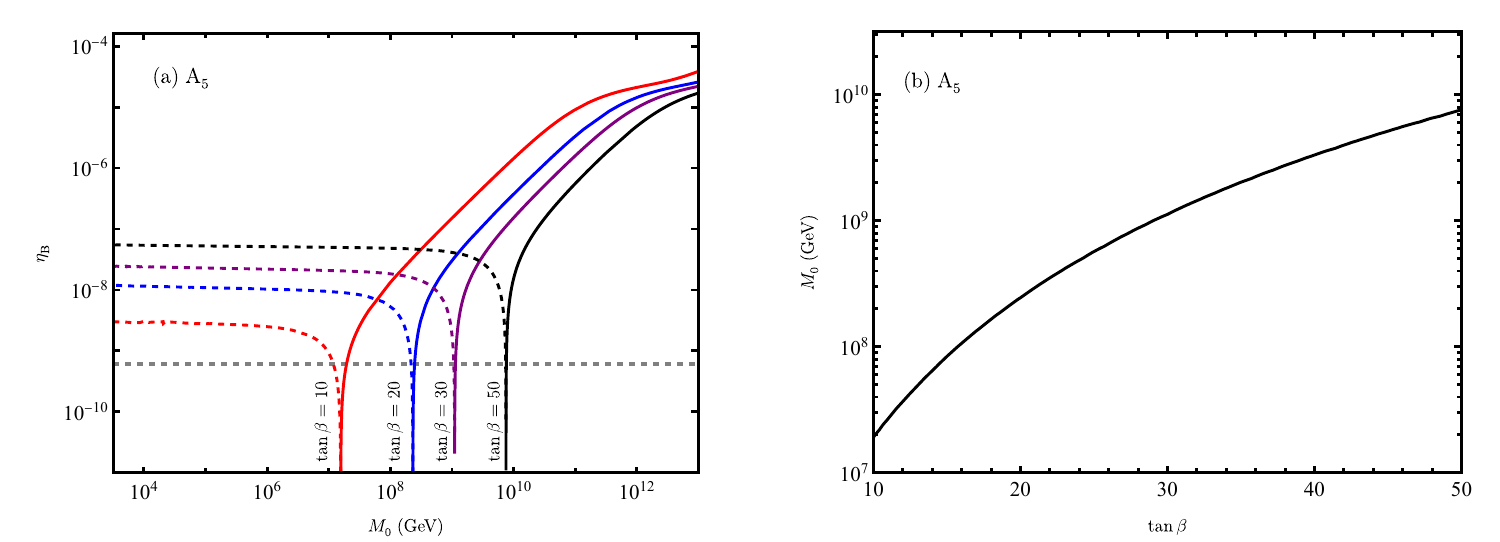}
	\caption{ For the modular ${\rm A}^{}_5$ symmetry model in section~6, the resulting $\eta^{}_{\rm B}$ as functions of $M^{}_0$ for the benchmark values of $\tan \beta =$10, 20, 30 and 50 (left panel). The horizontal gray dashed line shows the observed value of $\eta^{}_{\rm B}$. the values of $M^{}_0$ versus $\tan \beta $ that allow for a successful reproduction of the observed value of $\eta^{}_{\rm B}$ (right panel).}
	\label{fig13}
\end{figure}

Now, we study the RGE induced tri-resonant leptogenesis for the above model.
Figure~\ref{fig12} has shown the resulting CP asymmetries for the decays of $N^{}_I$ as functions of $M^{}_0$ for the benchmark values of $\tan \beta = 10$ and 30.
For $M^{}_0 < 10^{12}$ GeV, all of the CP asymmetries are almost independent of $M^{}_0$, while they change very sharply for $M^{}_0 > 10^{12}$ GeV.
The left panel of Figure~\ref{fig13} has shown the resulting $\eta^{}_{\rm B}$ as functions of $M^{}_0$ for some benchmark values of $\tan \beta$.
The result shows that the resulting $\eta^{}_{\rm B}$ is almost independent of $M^{}_0$ when $M^{}_0$ takes low values, and varies rapidly when $M^{}_0$ becomes sufficiently large.
Finally, the right panel of Figure~\ref{fig13} has shown the values of $M^{}_0$ versus $\tan \beta $ that allow for a successful reproduction of the observed value of $\eta^{}_{\rm B}$. The results show that the allowed values of $M^{}_0$ can vary between $10^7$ and $10^{10}$ GeV for $\tan \beta$ from 10 to 50.

 Before proceeding to the summary section, we briefly discuss the potential constraints imposed by charged lepton violating processes on the models considered in this work.
Among such processes, the most stringent limit stems from the $\mu \to e \gamma$ decay, as reported in Ref.~\cite{MEGII:2023ltw}:
	\begin{eqnarray}
		{\mathrm Br} \left( \mu \to e \gamma \right) < 3.1 \times 10^{-13}  \left( 90\% \;{\mathrm{C.L.}} \right)\;.
	\end{eqnarray}
Within the type-I seesaw framework, the branching ratio for $\mu \to e \gamma$ is given by~\cite{Ibarra:2011xn}:
\begin{eqnarray}
	{\mathrm Br} \left( \mu \to e \gamma \right) = \frac{2 \alpha^{}_{\rm em}}{32\pi}
	\left| \sum^{3}_{i=1} \left( R V \right)^*_{\mu i} \left( R V \right)^{}_{e i} \left[ G\left( X \right) - G\left(0\right) \right]\right|^2 \;,
\end{eqnarray}
where $X\equiv (M^{}_0/M^{}_W)^2$ with $M^{}_{1,2,3} \simeq M^{}_0$, and $G\left( X \right)$ is a loop integration function ranging in value from 4/3 to 10/3. Here,  $V$ denotes the unitary matrix that diagonalizes the Majorana mass matrix of heavy right-handed neutrinos, and $R$ is determined by $R^* \simeq M^{}_{\rm D} M^{-1}$.
This branching ratio can be approximated as:
\begin{eqnarray}
{\mathrm Br} \left( \mu \to e \gamma \right) \simeq \frac{2 \alpha^{}_{\rm em}}{32\pi}  \left( \frac{m^{}_{\nu} }{M^{}_0} \right)^2 \;.
\end{eqnarray}
For the heavy neutrino mass range considered in this paper ($10^3$ to $10^{15}$ GeV), the branching ratio lies on the order of $10^{-51}$ to $10^{-27}$, which is far below the $10^{-13}$ upper limit from the latest MEG II experiment.

\section{Summary}

Due to the peculiarity of the observed neutrino mixing pattern, there have been intensive model-building studies on neutrino-related physics with the employment of certain discrete non-Abelian symmetries. Particularly, in recent years, the idea of modular invariance has been proposed and become increasingly popular.

However, for some typical neutrino flavor-symmetry models, the right-handed neutrino mass matrix takes a form as shown in Eq.~(\ref{6}) which gives three exactly degenerate right-handed neutrino masses and consequently prohibits the leptogenesis mechanism to work successfully (to be specific, there would no CP asymmetries for the decays of right-handed neutrinos). But this kind of models can serve as a unique basis for realizations of the tri-resonant leptogenesis scenario: if the right-handed neutrino masses receive certain corrections so that their degeneracy is lifted to an appropriate extent, then the tri-resonant leptogenesis scenario will be naturally realized.

In this paper, we have considered four representative neutrino flavor-symmetry models which contain three exactly degenerate right-handed neutrinos, including a neutrino flavor-symmetry model realizing the TM1 mixing, the modular ${\rm A}^{}_4$ symmetry model, the modular ${\rm S}^{}_4$ symmetry model, and the modular ${\rm A}^{}_5$ symmetry model. For these models, we have performed a careful numerical calculation of the RGE induced mass splittings of right-handed neutrinos by employing the formulas as described in section~2.2 and subsequently the resulting baryon asymmetry by employing the formulas as described in section~2.1. The results show that, for appropriate values of the free parameters $M^{}_0$ and $\tan \beta$, the observed value of $\eta^{}_{\rm B}$ has chance to be successfully reproduced.
To be specific, we reach the following conclusions for the considered representative models:
\begin{itemize}
\item For the TM1 model in section~3, the observed value of $\eta^{}_{\rm B}$ can be successfully reproduced for $M^{}_0 \sim 7 \times 10^{12}$ GeV.
\item For the modular ${\rm A}^{}_4$ symmetry model in section~4, in the NO case the observed value of $\eta^{}_{\rm B}$ can be successfully reproduced when $M^{}_0$ is approximately between $10^{12}$ GeV and  $10^{13}$ GeV. While in the IO case the observed value of $\eta^{}_{\rm B}$ can not be reproduced successfully since the magnitudes of CP asymmetries are too large.
\item For the modular ${\rm S}^{}_4$ symmetry model in section~5,
for Case A and Case C there is an upper bound on $\tan \beta$ (about 45 and 28, respectively) and for each allowed value of $\tan \beta$ there may exist two allowed values of $M^{}_0$.
Importantly, Case C allows relatively low values of $M^{}_0$, which may be reachable for future direct measurements. For the other cases, the allowed values of $M^{}_0$ (about $10^{14}$, $10^9$ and $10^{13}$ GeV for Case B, Case D and Case E, respectively) are almost independent of $\tan \beta$.
\item For the modular ${\rm A}^{}_5$ symmetry model in section~6, the observed value of $\eta^{}_{\rm B}$ can be successfully reproduced for $M^{}_0$ in the range between $10^7$ and $10^{10}$ GeV for $\tan \beta$ from 10 to 50.
\end{itemize}

 For the four models discussed above, the energy scales required for successful leptogenesis are relatively high. This appears to contradict the common expectation that the resonant leptogenesis mechanism can lower the energy scale for successful leptogenesis to around the TeV scale. The reason lies in the distinct parameter constraints between general models and the ones studied here: In typical models, parameters are either unconstrained or subject to only loose restrictions, allowing resonant leptogenesis to effectively reduce the required energy scale to TeV or nearby values for successful leptogenesis. In contrast, the flavor structures of the models considered in this paper are tightly constrained by the imposed flavor symmetries which result in fewer free parameters (in fact, we are only left with the right-handed neutrino mass scale and $\tan \beta$ as free parameters) and thus stronger predictive power. This strong constraint has critical implications for the specific scenario of leptogenesis triggered by renormalization group running explored in our work: For these models, the TeV scale no longer accommodates successful leptogenesis. In fact, for some of them (see, e.g., Figures~9 and 12), a TeV-scale setup would lead to a lepton asymmetry larger than the observed value.

\vspace{0.5cm}

\underline{Acknowledgments} \vspace{0.2cm}

Zhi-Cheng Liu acknowledges the support from the Ph. D. Research Start-up Fund of Anyang Institute of Technology (No.~BSJ2021048).
Zhen-hua Zhao acknowledges the support from the National Natural Science Foundation of China under Grant No.~12475112, Liaoning Revitalization Talents Program under Grant No.~XLYC2403152, and the Basic Research Business Fees for Universities in Liaoning Province under Grant No.~LJ212410165050.

\end{document}